\documentclass[acmsmall,screen]{acmart}

\usepackage{graphicx}
\usepackage{textcomp}
\usepackage{xcolor}
\usepackage{enumitem}
\usepackage{tabularx}
\usepackage{booktabs}
\usepackage{multirow}
\usepackage{svg}
\usepackage{multicol}
\usepackage{adjustbox}
\usepackage{url}
\usepackage{todonotes} % Include the todonotes package
\usepackage{booktabs} % For better table lines
\usepackage{siunitx} % For unit and number handling

\AtBeginDocument{%
  }

\setcopyright{acmlicensed}
\copyrightyear{2024}
\acmYear{2024}
\acmDOI{XXXXXXX.XXXXXXX}

\acmJournal{JACM}
\acmVolume{37}
\acmNumber{4}
\acmArticle{111}
\acmMonth{8}

\begin{document}

\title{Unleashing OpenTitan’s Potential: a Silicon-Ready Embedded Secure Element for Root of Trust and Cryptographic Offloading}

\author{Maicol Ciani}
\email{maicol.ciani@unibo.it}
\orcid{0009-0003-7861-9129}
\affiliation{
  \institution{University of Bologna}
  \streetaddress{Viale del Risorgimento, 2}
  \city{Bologna}
  \country{Italy}
}
\author{Emanuele Parisi}
\email{emanuele.parisi@unibo.it}
\orcid{0000-0001-6607-7367}
\affiliation{
  \institution{University of Bologna}
  \streetaddress{Viale del Risorgimento, 2}
  \city{Bologna}
  \country{Italy}
}
\author{Alberto Musa}
\email{alberto.musa@unibo.it}
\orcid{0009-0003-1912-3801}
\affiliation{
  \institution{University of Bologna}
  \streetaddress{Viale del Risorgimento, 2}
  \city{Bologna}
  \country{Italy}
}
\author{Francesco Barchi}
\email{francesco.barchi@unibo.it}
\orcid{0000-0001-5155-6883}
\affiliation{
  \institution{University of Bologna}
  \streetaddress{Viale del Risorgimento, 2}
  \city{Bologna}
  \country{Italy}
}
\author{Andrea Bartolini}
\email{a.bartolini@unibo.it}
\orcid{0000-0002-1148-2450}
\affiliation{
  \institution{University of Bologna}
  \streetaddress{Viale del Risorgimento, 2}
  \city{Bologna}
  \country{Italy}
}
\author{Ari Kulmala}
\email{ari.kulmala@tii.ae}
\orcid{0009-0005-5755-7402}
\affiliation{
  \institution{Technology Innovation Institute}
  \streetaddress{P.O.Box: 9639, Masdar City}
  \city{Abu Dhabi}
  \country{United Arab Emirates}
}
\author{Rafail Psiakis}
\email{rafail.psiakis@tii.ae}
\orcid{0000-0001-6295-5476}
\affiliation{
  \institution{Technology Innovation Institute}
  \streetaddress{P.O.Box: 9639, Masdar City}
  \city{Abu Dhabi}
  \country{United Arab Emirates}
}
\author{Angelo Garofalo}
\email{angelo.garofalo@unibo.it}
\orcid{0000-0002-7495-6895}
\affiliation{
  \institution{University of Bologna}
  \streetaddress{Viale del Risorgimento, 2}
  \city{Bologna}
  \country{Italy}
}
\author{Andrea Acquaviva}
\email{andrea.acquaviva@unibo.it}
\orcid{0000-0002-7323-759X}
\affiliation{
  \institution{University of Bologna}
  \streetaddress{Viale del Risorgimento, 2}
  \city{Bologna}
  \country{Italy}
}
\author{Davide Rossi}
\email{davide.rossi@unibo.it}
\orcid{0000-0002-0651-5393}
\affiliation{
  \institution{University of Bologna}
  \streetaddress{Viale del Risorgimento, 2}
  \city{Bologna}
  \country{Italy}
}
%ORCIDs
%Emanuele Parisi \orcid{0000-0001-6607-7367}
%Alberto Musa \orcid{0009-0003-1912-3801}
%Maicol Ciani \orcid{0009-0003-7861-9129}
%Francesco Barchi \orcid{0000-0001-5155-6883}
%Davide Rossi \orcid{0000-0002-0651-5393}
%Andrea Bartolini \orcid{0000-0002-1148-2450}
%Andrea Acquaviva \orcid{0000-0002-7323-759X}

\renewcommand{\shortauthors}{Ciani et al.}




\received{03 March 2024}
%%\received[revised]{12 March 2009}
%%\received[accepted]{5 June 2009}

\begin{abstract}  
    The rapid advancement and exploration of open-hardware RISC-V platforms are catalyzing substantial changes across critical sectors, including autonomous vehicles, smart-city infrastructure, and medical devices. 
    Within this technological evolution, OpenTitan emerges as a groundbreaking open-source RISC-V design, renowned for its comprehensive security toolkit and role as a standalone system-on-chip (SoC). OpenTitan encompasses different SoC implementations such as Earl Grey~\footnote{\url{https://opentitan.org/book/hw/top_earlgrey/doc/datasheet.html}}, fully implemented and silicon proven, and Darjeeling~\footnote{\url{https://opentitan.org/book/hw/top_darjeeling/doc/datasheet.html}}, announced but not yet fully implemented.
    The former targets a stand-alone system-on-chip implementation, the latter oriented towards an integrable implementation. Therefore, the literature currently lacks of a silicon-ready embedded implementation of an open-source Root of Trust, despite the effort put by lowRISC on the Darjeeling implementation of OpenTitan.
    We address the limitations of existing implementations, focusing on optimizing data transfer latency between memory and cryptographic accelerators to prevent under-utilization and ensure efficient task acceleration. Our contributions include a comprehensive methodology for integrating custom extensions and IPs into the Earl Grey architecture, architectural enhancements for system-level integration, support for varied boot modes, and improved data movement across the platform. 
    These advancements facilitate the deployment of OpenTitan in broader SoCs, even in scenarios lacking specific technology-dependent IPs, providing a deployment-ready research vehicle for the community.
    We integrated the extended Earl Grey architecture into a reference architecture in 22nm FDX technology node, and then we benchmarked the enhanced architecture's performance analyzing the latency introduced by the external memory hierarchic levels, presenting significant improvements in cryptographic processing speed, achieving up to 2.7$x$ speedup for SHA-256/HMAC and 1.6$x$ for AES accelerators, compared to baseline Earl Grey architecture.
    
    % Addressing this gap, our study re-imagines OpenTitan beyond its original confines, proposing its transformation into a modular, integration-ready IP. 
    % This strategic pivot not only extends OpenTitan's applicability across diverse SoC designs but also optimizes its performance through targeted hardware extensions. 
    % By redefining OpenTitan as an embeddable IP, this research endeavors to fully exploit its security capabilities, significantly enhancing the security and integrity of future digital systems and marking a pivotal step toward realizing the full potential of open-hardware RISC-V platforms in the semiconductor ecosystem.
\end{abstract}

\maketitle

\section{Introduction}

The move towards open-hardware RISC-V platforms is shaping the future of edge computing across critical sectors such as aerospace~\cite{aeropspace}, automotive~\cite{automotive}, and healthcare devices~\cite{medical}.
These areas increasingly rely on heterogeneous architectures~\cite{fulmine} composed of Linux-capable hosts, providing software portability and ease of development for user applications, coupled with various accelerators dedicated to efficiently running data-intensive workloads at the edge. 
As these platforms are often employed in safety-critical fields, they must be secure and reliable, guaranteeing security properties such as sensitive data, code protection, and software execution integrity and providing means to run standard platform-level security features efficiently, such as secure boot. 
In this context, silicon Root-of-Trusts~\cite{RoT} (RoT) represents the state-of-the-art in terms of trusted computing and system integrity, establishing an isolated silicon region with security features for data and code protection, such as tamper-proof memory and cryptographic accelerators, and physical countermeasures, such as voltage and temperature monitoring, aiming at detecting potential security threats.

An example of state-of-the-art RoT is OpenTitan~\footnote{\url{https://opentitan.org/}}, released by lowRISC as the first open-source RISC-V design that introduces a silicon Root-of-Trust SoC. 
OpenTitan is equipped with a set of cryptographic accelerators designed to safeguard data confidentiality, integrity, and authenticity, supporting the computation of hash functions (SHA-256 and SHA-3), message authentication codes (HMAC and KMAC), and symmetric encryption such as Rivest Shamir Adleman (RSA), elliptic curve cryptography (ECC) with the most common modes of operations. 
Additionally, the acceleration of integer operations on large numbers is supported OpenTitan Big Number accelerator (OTBN), boosting the system's performance when asymmetric encryption schemes are used to implement digital signature or key exchange protocols.
OpenTitan includes advanced features such as memory data and address scrambling, complemented by Error Correction Codes (ECC) for enhanced data integrity.
It is provided with a secure one-time-programmable (OTP) memory based on eFuse to permanently store the lifecycle states and cryptographic seeds and an entropy source based on Physical Unclonable Functions (PUF).

OpenTitan encompasses different SoC implementations such as Earl Grey~\footnote{\url{https://opentitan.org/book/hw/top_earlgrey/doc/datasheet.html}} and Darjeeling~\footnote{\url{https://opentitan.org/book/hw/top_darjeeling/doc/datasheet.html}}.
While Earl Grey is designed as a stand-alone SoC, Darjeeling's specification is oriented towards an embeddable implementation of OpenTitan.
The literature currently lacks an embedded implementation of an open-source Root of Trust, despite the effort put by lowRISC on the Darjeeling implementation of OpenTitan, as the latter has been announced but is still under development and has yet to be released.
On the other hand, Earl Grey's design was taped out by lowRISC in 2023, and it is provided with a comprehensive verification framework to emphasize security, reliability, and functionality through formal methods, extensive simulation testing, continuous integration, FPGA prototyping, and audits to ensure the integrity and security of its components. 
For this reason, Earl Grey represents the ideal base architecture to be extended and customized for an embeddable use case.
Moreover, previous studies~\cite{parisi2024assessing} highlight that the performance of OpenTitan cryptographic accelerators is bounded by data transfer latency between integrated memories and accelerators. 
This may lead to an under-utilization of the cryptographic accelerators' bandwidth if the datapath between memory and accelerator needs to be optimized. As a result, the performance of OpenTitan as an on-chip security-hardened hub for the acceleration of security tasks may be jeopardized.

Building upon the foundational work of OpenTitan, we extend the Earl Grey architecture to turn it into a silicon-secure element easily integrable within larger designs.
We also facilitate its usage as an extendable research platform and make integrating novel features easier.
For this purpose, we present a framework to extract from the Earl Grey SoC's digital system architecture in charge of handling the security features and integrate it into broader SoCs, considering different scenarios where the technology-dependent Intellectual Properties (IPs) of embedded flash, eFuse, and PUF are unavailable for the target technology node, enhancing OpenTitan's versatility as a research vehicle.
Furthermore, we extended OpenTitan datapath to optimize data movement towards the accelerators, customizing the TLUL interconnect and integrating a Direct Memory Access (DMA) engine along with a Tightly Coupled Data Memory (TCDM) dedicated exclusively to the accelerators.
We present a comprehensive methodology for OpenTitan deployment on FPGA and ASIC.
Finally, we present the implementation of the enhanced OpenTitan design in a technology node, and we benchmark its performance as a cryptographic accelerator, comparing it with the default architecture released by lowRISC. We propose an extension of the work presented by Parisi et al. \cite{parisi2024assessing}, which introduces the performance assessment of the Open Titan architecture as an accelerator for cryptographic functions, with the following main contributions:
\begin{itemize}
    \item A methodology to extend the OpenTitan Earl Grey architecture with custom extensions and IPs, remaining compatible with the lowRISC design flow, enabling easy system upgrading (e.g., in case of bug fixes or extensions from lowRISC). The proposed approach is generic and can be applied to Darjeeling once the design is finalized by lowRISC. 
    \item Architectural extensions to enable Earl Grey system-level integration into broader SoCs, including designing a mailbox-based communication mechanism to enable offloading tasks from a host processor to the embedded secure element.
    \item Architectural extensions to support different boot modes, facilitating silicon prototyping of embedded secure elements based on OpenTitan architectures in a context where technology-dependent IPs such as eFuse, embedded flash, anti-tamper sensors, or PUF might not be available to researchers.
    \item A benchmarking of SHA-256, HMAC, and OTBN cryptography algorithms and datapath extensions to enhance data movement across the platform, increasing the encryption performance of the cryptographic engines.
\end{itemize}

We demonstrate the proposed methodology and architectural extensions implementing in Global Foundries 22nm FDX technology node the reference heterogeneous SoC with the extended Earl Grey architecture integrated, resulting in a total silicon area of 7.28$mm^2$, with the embedded secure element accounting for 1.7$mm^2$, about 30\% of the area. 
The maximum speed-up we observe with respect to the corresponding pure software implementation of the benchmarks are 2.7$x$ and 1.6$x$ for SHA-256/HMAC and OTBN, respectively. 
In this scenario, we maximized the performance of the cryptographic accelerators, saturating their nominal bandwidth when working with payloads stored within OpenTitan's perimeter.

% OpenTitan is designed as a flexible and configurable platform. The interconnect and the memory configuration can be changed with minimal effort. 
% However, OpenTitan is a system-on-chip and the core architecture responsible for managing the security model is not a standalone embedded secure element.
% Its inputs are intricately linked to the rest of the SoC, lacking a flexible input/output interface crucial for system-level integration in broader SoCs.
\section{Related Work}
%\missing{first two sentences can be removed if needed. Some references on SW RoT and reference that assess that SWRoT are less effective than HWRoT}
%A trusted foundation is paramount in modern Cyber-Physical Systems (CPS) as it  ensures the integrity, confidentiality, and availability of these systems, safeguarding against malicious exploits and unauthorized access. 
%At the same time, also executing cryptographic workloads with high performance is a concern especially for networking applications. 
%In this section we review existing solutions that ensure integrity properties to the compute systems and that accelerate cryptographic functions for networking.
A trusted foundation is paramount in modern Cyber-Physical Systems (CPS) as it ensures execution integrity, data confidentiality, and system availability, safeguarding against malicious parties exploiting bugs and vulnerabilities to gain unauthorized access to the system.
Implementing and deploying security protocols are often based on executing cryptographic primitives to ensure data integrity or communication confidentiality and authentication.
Executing such workloads with high performance is essential to ensure security does not become a system bottleneck, and the designer never needs to face the security versus performance trade-off.
The following subsections describe related work about hardware-assisted trusted execution environments for code and data integrity and confidentiality, as well as existing approaches to optimizing the execution of cryptographic functions and protocols in modern CPS.
Finally, we summarize state-of-the-art research works that have exploited, characterized, or enhanced the OpenTitan Root-of-Trust for hardening RISC-V platform's security properties.

%
%RoT can fall in three categories: software-based RoT (SWRoT), hardware-based RoT (HWRoT) and hybrid solutions. 
%
%Despite the cost-effectiveness, flexibility and ease of updates of SWRoT, they are less effective than HWRoT against software attacks and they provide limited physical protection, expanding the attack surface. Our review is therefore more focused on HWRoT or hybrid solutions. 

\subsection{Hardware-assisted trusted execution environments (TEEs)}
%\missing{to give a pass to the text. Eventually some parts can be shortened}

%
To ensure the secure and reliable execution of applications in potentially insecure environments, it is required to create secure enclaves within the device where the code can execute safely and data is protected. One solution to achieve this goal is through  hardware-assisted trusted execution environments (TEEs).

Intel SGX~\cite{zheng2021survey} (Software Guard Extensions) is a technology developed by Intel that enables the creation of secure enclaves within the CPU, through custom extensions to the instruction set of the processor's core. SGX provides a hardware-based solution for secure execution of code and protection against both physical and software attacks, and is widely adopted in cloud computing environments. %However, recent works~\cite{nilsson2020survey} have raised potential risks on side-channel attacks.

ARM TrustZone~\cite{ngabonziza2016trustzone, pinto2019demystifying} is a security extension for ARM processors, it provides features such as secure boot and cryptographic operations, enabling the creation of secure and non-secure execution environments on a single chip. It is commonly used in, but limited to, ARM-based mobile devices and embedded systems to establish a secure foundation. 

With the ever-increasing spread of the RISC-V Instruction Set Architecture (ISA), many research efforts aim to improve security solutions on top of RISC-V-based processors as well~\cite{ehret2020hardware,kumar2020towards}. Hector-V~\cite{nasahl2021hector} and SiFive WorldGuard~\cite{sifive-worldguard} are two examples that provide hardware-assisted solutions to guarantee strong isolation properties. These are essentials to build TEEs, such as SANCTUM~\cite{costan2016sanctum} and Keystone~\cite{lee2020keystone}, that provide secure enclaves for executing sensitive tasks.

However, such solutions are not primarily designed to serve as a root of trust (RoT) for the entire system; instead, they focus on protecting specific code and data within secure enclaves. Therefore, they might be exposed to side-channel and other physical attacks~\cite{nilsson2020survey}. Unlike TEEs, Hardware Root of Trusts (HWRoTs) establish a secure foundation for the entire system, incorporating robust cryptographic measures and attestation mechanisms, offering superior protection against various threats, including side-channel attacks.

Trusted Platform Module (TPM)~\cite{module2018trusted} is a widely adopted HWRoT standard, especially in PCs and servers; it was defined by the Trusted Computing Group and standardized by ISO in 2009. TPM specifies a dedicated microcontroller that provides a secure foundation for cryptographic functions, including key generation, storage, and attestation. Although it provides a secure storage and execution environment for cryptographic keys, it shows limited processing capabilities for advanced crypto functions. It does not cover all security concerns of modern CPS. Hardware Security Modules (HSMs)~\cite{sommerhalder2023hardware} are specialized hardware devices that provide similar functionalities (secure key storage, cryptographic functions) and are used in various applications and services, such as secure key storage, digital signatures, and transaction security. They are expensive solutions compared to others and often external to the target devices.
Another solution is Pluton~\cite{stiles2019hardware}, a security processor developed by Microsoft in collaboration with AMD, Intel, and Qualcomm. It is designed to enhance the security of Windows PCs by providing a dedicated chip for key management and other security functions. Similar solutions that target also the protection of embedded devices are the ATECC608A secure element by Microchip~\cite{microchip-rot}, the OPTIGA TPM by Infineon~\cite{optiga-infineon}, the RT-600 Series HWRoT~\cite{rambus-rt600} by Rambus, or contributions from academia~\cite{aliaj2021garota, gui2018hardware}.

Targeting embedded systems and mobiles, the security chip Titan M by Google used in Pixel smartphones provides a secure enclave for handling sensitive operations such as cryptographic processing, device verification, and secure boot, enabling the implementation of a Trusted Execution Environment (TEE) in Tamper Resistant Hardware. Additionally, for the first time, it introduces two new features to TPMs that did not exist before first-instruction integrity and remediation property~\cite{frazelle2020securing}. The first one allows the verification of the earliest code that runs on each machine's startup cycle, guaranteeing the integrity of the code at boot time. The second one securely allows the update of the secure firmware of the chip, even after the manufacturing process; this enables higher flexibility to fix bugs or update the firmware's features. 

The mentioned Hardware Root of Trust (HWRoT) solutions are typically presented as standalone System-on-Chip (SoC) units integrated into the computing system's motherboard for protection and communication through I/O peripherals. In addition, they may be proprietary solutions closely tied to specific processors or exclusive vendor offerings. 

These traits restrict their versatility across a broader range of devices and hinder their use for research purposes. One solution towards developing a re-usable RoT silicon level IP is given by the Open Compute Project (OCP) Community that recently presented Caliptra~\cite{caliptra_prj}. It targets cloud computing platforms to protect vendor-agnostic CPUs, GPUs, and SSDs. The implementation is not available despite the specifications being released open-source~\footnote{https://github.com/chipsalliance/Caliptra}. A fully open-source RoT solution, targeting cloud but also edge and smaller systems, is given by LowRisc, with the Open Titan project.

Inspired by the Titan M chip of Google, Open Titan is the first open-source silicon root of trust that aims to provide an open-source, transparently designed, and high-quality silicon Root of Trust (RoT) chip. 
The purpose of OpenTitan is to create a secure and flexible foundation for building trustworthy hardware systems by open-sourcing the design and making it freely available~\footnote{https://github.com/lowRISC/opentitan}, thus increasing trust and transparency in the semiconductor industry and enhancing the security of hardware platforms deployed in security-sensitive scenarios, spanning from autonomous vehicles, to healthcare implantable devices and critical industrial infrastructure.
\subsection{Open Titan Project}

The OpenTitan RoT has recently received significant attention in the security research community, and many literature works have been published exploring various aspects of its design, security properties, and performance characteristics.~\cite{post_quantum,antognazza2021metis,parisi2024titancfi,parisi2024assessing,ciani2023cyber}.

Meza et al.~\cite{meza2023security} apply information flow tracking (IFT) security verification flow to perform hardware security verification of the Open Titan One Time Programmable (OTP) controller, that plays a crucial role in the overall security of the system.  The authors identified a weakness in the controller, localized the source of the error, and developed a hardware patch that was accepted as a pull request into the open-source OpenTitan repository.

Authors of ~\cite{post_quantum} enable lattice-based post-quantum cryptography in Open Titan by providing dedicated extensions to the ISA of the OpenTitan Big Number Accelerator (OTBN). 
The solution proposed comply with the OpenTitan adversary model and PQC performance and security level have been tested in digital signature verification applications, showing significant improvement compared to the baseline OpenTitan design.

Parisi et al.~\cite{parisi2024titancfi} developed a novel methodology for implementing custom Control-Flow Integrity (CFI) policies in OpenTitan firmware, harnessing the RoT as a security co-processor to protect platform host cores against control flow violations. 
Leveraging secure, tamper-proof memory within the RoT, their approach mitigates the risk of exposing security-sensitive metadata, while employing RoT security accelerators to authenticate control-flow sensitive metadata spilled to platform memory.

The authors of~\cite{parisi2024assessing} performed a characterization of the OpenTitan performance when used as cryptographic accelerator, identifying key architectural bottlenecks which prevent the OpenTitan firmware to fully leverage the bandwidth guaranteed by the available OpenTitan accelerators.

Furthermore, Ciani et al.~\cite{ciani2023cyber} explored novel security policies and protocols deployed on top of the OpenTitan platform, showcasing a use case where OpenTitan detects a security breach on the SoC aboard the MAV and drives its exclusive GPIOs to setup an unconventional visual communication between two palm sized MAVs.

However, despite these advancements, notable gaps persist in our understanding of OpenTitan's architectural constraints, particularly concerning its integration within heterogeneous CPS architectures. 
Moreover, the absence of research addressing scenarios devoid of technology-dependent IPs like eFuses and on-chip FLASH memories poses challenges for broader academic explorations. 
While initiatives like the Darjeeling project~\cite{darjeeling} have emerged to address some of these limitations, they remain in nascent stages, leaving room for further enhancements. 
In light of these considerations, our work aims to extend OpenTitan to serve as a technology-independent SoC-integrated RoT research framework, bridging existing gaps and fostering innovation in secure hardware design. 
Through collaborative efforts, we envision OpenTitan evolving into a cornerstone of trust and resilience in the burgeoning landscape of embedded systems security.

% Very relevant work for us as well
%https://www.researchgate.net/publication/346533633_Towards_Designing_a_Secure_RISC-V_System-on-Chip_ITUS
% file:///C:/Users/redar/Downloads/Kumar2020_Article_TowardsDesigningASecureRISC-VS.pdf

%\subsection{Acceleration of Cryptography functions for networking}
\subsection{Acceleration of Cryptography Operations}

The implementation and deployment of security protocols in modern systems, particularly in the context of Internet of Things (IoT) devices, rely cryptographic primitives to ensure data integrity, communication confidentiality, and authentication. 
However, ensuring high performance while executing such cryptographic workloads is crucial to prevent security from becoming a bottleneck in the system and to eliminate the need for designers to compromise between security and performance.

In IoT devices, optimization of cryptographic workloads is typically achieved through the integration of accelerators within the system-on-chip~\cite{wang2021energy,tehrani2020riscv,pham2020implementing}. 
These accelerators optimize common cryptographic tasks and can be used either as standalone memory-mapped IPs or incorporated within the main host core,and accessed by extending the core ISA.
Recent research has proposed various solutions to enhance the efficiency of cryptographic operations in IoT devices. 

Wang et al.~\cite{wang2021energy} introduced an energy-efficient crypto-coprocessor that supports essential cryptography algorithms like AES, ECC, and SHA. 
This design, integrated within the RISC-V Rocket chip core, 
acts as a crypto-extension activated by custom instructions through the Rocket Coprocessor Interface.
Similarly, Tehrani et al.~\cite{tehrani2020riscv} have contributed by adding support for lightweight cryptographic algorithms tailored for fast encryption on IoT devices, while Marshall et al.~\cite{pham2020implementing} have implemented the draft of the official RISC-V ISA extension for scalar cryptography. 
These approaches alleviate the burden of executing cryptographic tasks in software, thereby enhancing overall system performance.

Despite these advancements, certain fundamental security challenges persist since data remains vulnerable to unauthorized access in memory, and encryption keys are stored in clear within general-purpose memory. 
Furthermore, there is a lack of protection against side-channel attacks and physical intrusion.
Addressing these criticalities, HECTOR-V~\cite{nasahl2021hector} proposes a TEE where security-critical tasks are offloaded to a security-hardened core featuring sponge-based execution integrity~\cite{werner2018spongebased}. 
Additionally, HECTOR-V introduces the concept of trusted I/O path, allowing secure sharing of devices on the SoC by assigning unique identifiers to each system component and implementing mechanisms for secure device access control.

In this context the utilization of OpenTitan as a secure accelerator for cryptographic workloads pose a promising avenue for enhancing security while maintaining performance.
OpenTitan offers dedicated hardware accelerators for cryptographic tasks, thus addressing security concerns while providing efficient cryptographic operations, while featuring special private storage for cryptographic keys and a set of countermeasures against side-channel attacks.
However, significant challenges remain in fully leveraging OpenTitan's capabilities. 
The native design provided by lowRISC lacks essential features such as Direct Memory Access for automatic data movement and features a high-latency private interconnect. 
These limitations hinder OpenTitan's ability to efficiently utilize its cryptographic accelerators, resulting in poor bandwidth usage.
Overall, while significant progress has been made in optimizing cryptographic workloads and enhancing security in IoT devices, 
further research and development are necessary to address remaining challenges and fully exploit the potential of emerging 
technologies like OpenTitan.

%\missing{Description of the three aforementioned classes of crypto functions (Fetch crypto descriptions from sections 2.4.x).}

% https://www.edacentrum.de/making-authentication-token-ic-based-open-titan-project --> I cannot find much on this...

% security verification of Open Titan - restuccia
% https://ieeexplore.ieee.org/stamp/stamp.jsp?arnumber=10106105

% Security module on FPGA
% https://dl.acm.org/doi/abs/10.1145/3600160.3605168

% Extension of Open Titan Against Side Channel Attacks
%https://ieeexplore.ieee.org/document/9424552

% Quantum Criptography on Open Titan. to look carefully, they intergrated an accelerator in the OCBN
%https://dl.acm.org/doi/pdf/10.1145/3605769.3623993

% Very relevant work for us as well
%https://www.researchgate.net/publication/346533633_Towards_Designing_a_Secure_RISC-V_System-on-Chip_ITUS
% file:///C:/Users/redar/Downloads/Kumar2020_Article_TowardsDesigningASecureRISC-VS.pdf

%Open Titan DarJeliing (obviously the closest work)
%https://opentitan.org/book/hw/top_darjeeling/doc/datasheet.html
\section {Background}
\label{sec:background}

\subsection{OpenTitan Project}

\begin{figure}[b!]
  \centering\includegraphics[width=0.85\columnwidth]{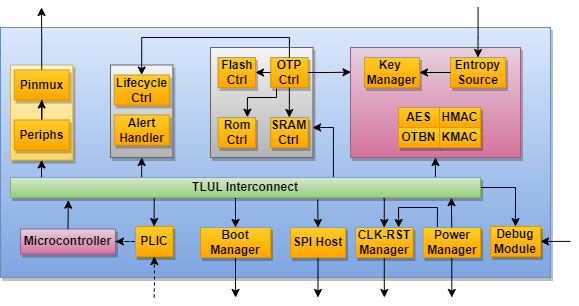}
  \caption{\texttt{Block diagram of top\_earlgrey} module, the core digital architecture of Earl Grey SoC.}
  \label{fig:ot-arch}
\end{figure}

OpenTitan is a security module available as an open-source system-on-chip design by lowRISC, serving to establish a secure enclave with a hardware RoT. 
The OpenTitan project provides different versions of the same architecture, such as Earl Grey and Darjeeling, targeting applications such as data centers, cloud systems, and embedded systems. 
OpenTitan protects sensitive data and systems against hardware attacks, tampering, and counterfeiting. 
The specific OpenTitan instance we consider in this study is the Earl Grey architecture, depicted in Figure \ref{fig:ot-arch}.
OpenTitan includes a secure microcontroller (Ibex core), on-board private memory, cryptographic hardware accelerators, and communication I/O interfaces.

The secure boot process in OpenTitan begins with the ROM stage for initial device setup and authentication of the next stage, ROM\_EXT, using public keys from Silicon Creator. 
The ROM stage initializes hardware and checks the ROM\_EXT manifest for authenticity and compatibility with the device's current state before proceeding.
The ROM\_EXT stage, stored in Flash, further authenticates the next boot stage, manages key transitions, and performs critical boot services.
The same framework can be applied to eventual further boot stages and can be extended to the external platform in an embedded use case.
This multi-stage approach guarantees a secure boot process by ensuring each component is authenticated by its predecessor, leveraging hardware features and cryptographic checks to protect against unauthorized modifications or executions.
% OpenTitan implements a secure boot process that permits trusted software to execute, ensuring the system's integrity at boot time. 
Cryptographic acceleration and memory protection mechanisms contribute to swift encryption and secure data storage. 

The hardware architecture centers around the Ibex core, a 32-bit RISC-V CPU designed to support embedded and power-efficient applications~\cite{schiavone2017slow}, optimized for minimal area usage. 
The system-level interconnect is based on the TileLink Uncached Lightweight\footnote{\url{https://static.dev.sifive.com/docs/tilelink/tilelink-spec-1.7-draft.pdf}} (TLUL) protocol. 
The memory domain of OpenTitan consists of a scratchpad SRAM memory and an embedded flash memory, each equipped with separate controllers, including Error Correcting Code (ECC) and data-address scrambling to enhance security and reliability. 
Essential cryptographic and scrambling keys are generated by accelerators and stored in a secure, tamper-proof region: the one-time-programmable eFuse memory. 
It incorporates a key manager responsible for managing hardware identities and root keys and protecting critical assets from software threats.
The security domain also includes a hardware life cycle controller, which dynamically enables or disables features based on the device's life cycle state.
OpenTitan features a peripheral subsystem supporting UART, USB, I2C, SPI, and GPIO for communication with the external world. 
Since Earl Grey is designed as a stand-alone SoC, some extensions are required to integrate it into broader SoCs. 
Remarkably, it is necessary to expose to the in-out interface a master port driven by the central TL-UL interconnect and a bridging module to convert the TL-UL protocol, adopted by OpenTitan, into the one featured by the host platform.

Dedicated hardware accelerators have been incorporated to handle cryptographic computations efficiently. 
These accelerators are specifically designed to support three essential cryptographic algorithms mandated by the security features targeted by the silicon Root-of-Trust, namely AES, SHA-256, SHA-3, HMAC and KMAC.
Moreover, the OTBN enables accelerating asymmetric encryption algorithms crucial to key exchange mechanisms and digital signature schemes.

\subsubsection{HMAC and SHA-256}
SHA-256 is a secure hashing function widely utilized to ensure data integrity in applications such as digital signatures and certificate generation. 
This algorithm processes an input message to produce a hash value of fixed size, employing a one-way design that prevents reverse engineering or finding different inputs leading to the same hash value. 
HMAC generates a secure digest using cryptographic hash functions, creating a message authentication code that depends on a secret key shared between the sender and recipient.
OpenTitan includes an HMAC accelerator programmed to calculate SHA-256 and HMAC message digests. 
The front-end of the accelerator comprises a memory-mapped FIFO where the software pushes the next 64-byte data block for processing. 
Once the FIFO is filled, the accelerator automatically reads it and initiates an 80-cycle procedure to update its internal state before processing the next data block.
The software sets up the accelerator in SHA-256 or HMAC mode, specifying the message and digest endianness. 
If configured in HMAC mode, the software loads the secret key into designated control registers. 
It then enters a loop where it waits for the FIFO to be empty before pushing the next 64-byte data block.

\subsubsection{AES}
The Advanced Encryption Standard (AES) is a symmetric encryption algorithm that operates on a 16-byte block, supporting key lengths of 128, 192, and 256 bits. 
The AES accelerator in the OpenTitan Root-of-Trust optimizes storage utilization through real-time generation of round keys, reducing storage needs and accelerating key-switching activities. 
Upon reset or when software activates, the device erases the primary key and data registers using pseudo-random data to reduce the possibility of side-channel leakage.
The front-end of the accelerator comprises a 16-byte FIFO where the software driver loads the plaintext. 
When configured in automatic mode, upon loading the input FIFO, the accelerator immediately encrypts its content and generates the next ciphertext block based on the chosen mode of operation. 
This work focuses on an OpenTitan design incorporating the default masked AES accelerator implementation and utilizes a CBC mode with a 256-bit key. 
Under these conditions, the encryption of a 16-byte plaintext block takes 72 cycles. 
The operational principles of this accelerator are similar to those of the HMAC accelerator. 
Once control registers are set up for the accelerator and both key and initialization vector are loaded, the software enters the processing loop within which it awaits completion of acceleration operations before retrieving ciphertext from output FIFO and feeding in subsequent plaintext blocks for processing into input FIFO.

\subsubsection{RSA and OTBN}
OpenTitan enhances the performance of mathematical operations in public-key cryptosystems by delegating them to the OTBN co-processor. 
This co-processor has a tailored processor for broad integer arithmetic, a 32-bit control path, and a 256-bit data path. 
The OTBN offers comprehensive support for control flow with conditional branches, unconditional jumps, and hardware loops. 
In addition, it implements custom instructions to support standard functions in public-key cryptosystems, like pseudo-module addition or partial-word multiplication and accumulation.
This work focuses on RSA, a commonly used form of asymmetric encryption in digital signature and key-exchange protocols. 
Its security is based on the challenge of factoring the product of two large prime numbers to undo the exponentiation operations required for message encryption or decryption.
Accelerating asymmetric encryption on the OTBN involves implementing the specific algorithm using the OTBN instruction set. 
Then, the Ibex secure microcontroller within OpenTitan loads the program binary and data into reserved memories in OTBN. 
Lastly, a command to start computation is activated on OTBN, with OpenTitan waiting for operation completion by checking a control register in the OTBN interface.

\subsection{Reference System Architecture}

\begin{figure}[t!]
  \centering\includegraphics[width=\columnwidth]{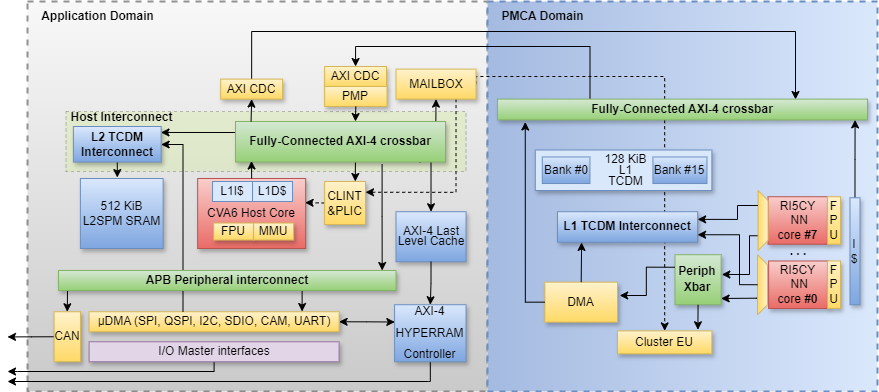}
  \caption{Block diagram of the reference heterogeneous architecture.}
  \label{fig:SoC_arch}
\end{figure}

The reference system for the contribution of this work, depicted in Figure \ref{fig:SoC_arch}, is an open-source heterogeneous architecture based on the Shaheen SoC~\cite{shaheen-TCAS-I}, and it targets a wide range of emerging mixed-criticality IoT applications, including industrial automation, robotics, automotive, space and autonomous UAV navigation.
The heterogeneous computing system is built around the linux-capable 64-bit CVA6~\cite{cva6} application processor. The CVA6 core is a six-stage pipeline RISC-V processor featuring advanced branch prediction and a comprehensive memory management unit (MMU) for hardware address translation. It provides separate TLBs for data and instructions, ensuring robust exception handling across all pipeline stages. Adhering to the RISC-V privileged specification, it supports atomic operations and manages execution modes through control status registers (CSRs). The hierarchical cache system offers configurable first level caches and an optional second level cache for optimal performance and power usage. 

The central on-chip communication happens on an AXI4-based bus. In contrast, the off-chip communication capabilities satisfy the target applications' domain requirements, providing a comprehensive set of peripherals reachable from the main AXI interconnect through an APB bridge. These include common UART, SPI, I2C, PWM protocols, and Ethernet, CAN, and camera interfaces. A DMA controller is incorporated to move data between the memory and peripherals efficiently without employing the application processor. 
The memory hierarchy consists of an instruction and data cache (L0) for each core, a platform-level embedded scratchpad memory (L2) supporting the light real-time requirements for many embedded applications, and a 512KB last-level cache (LLC) that refills on external memory (L3).
High-performance (LP)DDR controllers are typically employed in high-end heterogeneous systems to access an external DDR main memory. 
However, these components are power-hungry, extremely expensive, technology-dependent (requiring dedicated Phys that cannot be open-sourced due to links to the specific technology used for implementation), and, most importantly, unavailable to most universities and research institutes. 
For this reason, in our work, we employ a fully digital and fully open-source HyperRAM controller IP accessing the L3 DRAM, which can be fully implemented with foundation libraries available to University services such as Europractice~\cite{das2014europractice}, contributing to the widespread of the project to the open source community. 

Moreover, the heterogeneous architecture incorporates a Parallel multicore accelerator (PMCA) tailored for machine learning and digital signal processing. 
The system's PMCA revolves around eight CV32E-based RI5CY~\cite{riscy} processors designed explicitly for high energy-efficient digital signal processing in tightly coupled multicore clusters.
The hosting platform is equipped with hardware virtualization capabilities~\cite{h-extension} fully compliant with the RISC-V Hypervisor extension. 
This feature enables the secure simultaneous operation of a Real-Time Operating System (RTOS) and a comprehensive Operating System (OS) on the same host processor core. 
To further segregate the activities of these concurrent software environments (trusted and untrusted), avert security risks, and facilitate operations across multiple domains, the host processor incorporates Physical Memory Protection (PMP), along with ISA and micro-architecture enhancements for reducing timing attack vulnerabilities~\cite{nielsCva6}.
The primary aim of the proposed work is to extend the architecture presented by Valente et al. ~\cite{shaheen-TCAS-I} with an embedded secure element based on the OpenTitan Earl Grey architecture, integrating the presented security and accelerated cryptography functions.

\section{Open Titan as a Silicon-Ready Root of Trust Co-Processor for Security Management and Crypto Acceleration}

This section outlines our extensions implemented to the Earl Grey architecture for its seamless integration into larger SoCs as an embedded Root-of-Trust and cryptographic accelerator. 
First, we present our general methodology for deploying OpenTitan in a different design environment.
Second, we detail the architectural modifications to ensure compatibility with diverse SoC platforms.
Furthermore, we delineate modifications to Earl Grey's internal architecture to improve computational efficiency in cryptographic operations. 
Specifically, we address inefficiencies in data movements that hinder meeting the bandwidth requirements of the cryptographic accelerators. 
Our proposed optimizations include optimizing the accelerators' interfaces, designing a dedicated TCDM subsystem, and integrating an autonomous DMA engine for efficient data transfer across OpenTitan perimeter, from and towards the external memory hierarchy. 

\subsection{General Methodology}

lowRISC has enhanced OpenTitan's flexibility by developing an advanced configuration system. 
This system leverages a comprehensive suite of Python scripts, enabling the dynamic configuration and generation of Earl Grey's top level module, the interconnect, the Platform-Level Interrupt Controller (PLIC), and the flash controller. 
These scripts take a series of SystemVerilog templates and HJSON configuration files as inputs. 
These files delineate various design specifics, such as the IP components to be instantiated within Earl Grey's top level, IP interconnections, TLUL interconnects' ports, and the memory map. 
The configuration phase primarily utilizes three scripts: TLGEN for the tile link generator, IPGEN for IP block generation, and TOPGEN for assembling the top-level module. 
OpenTitan’s framework employs Bazel\footnote{\url{https://bazel.build/}} tool for handling hardware implementation targets, such as RTL simulations and FPGA emulation (employing Fuse-soc), and for building the software images (employing RISCV toolchain).
PULP's framework relies on the Mentor QuestaSim RTL simulator and Bender\footnote{\url{https://github.com/pulp-platform/bender}} dependencies manager.
QuestaSim provides more extensive features for in-depth simulation, including better support for system-level verification, mixed-signal simulation, and recording the core's execution traces.
On the other hand, Bender manages SystemVerilog source files and the various hardware implementation targets through manifests, similarly to Bazel.
The manifest includes all the RTL source files grouped in targets and external git dependencies. It provides utilities to generate TCL files for different hardware implementation's deployments.
For each target, Bender generates TCL scripts including the source files associated with that target.

Fuse-soc could accomplish the same objective as Bender, but it is not supported in PULP platform's environment, and consequently also in target SoC's one. Furthermore, Bender offers the possibility to handle different targets (simulation, synthesis, etc.) without changing the manifest or the scripts, while Fuse-soc cannot.
\begin{figure}[t!]
  \centering\includegraphics[width=0.80\columnwidth]{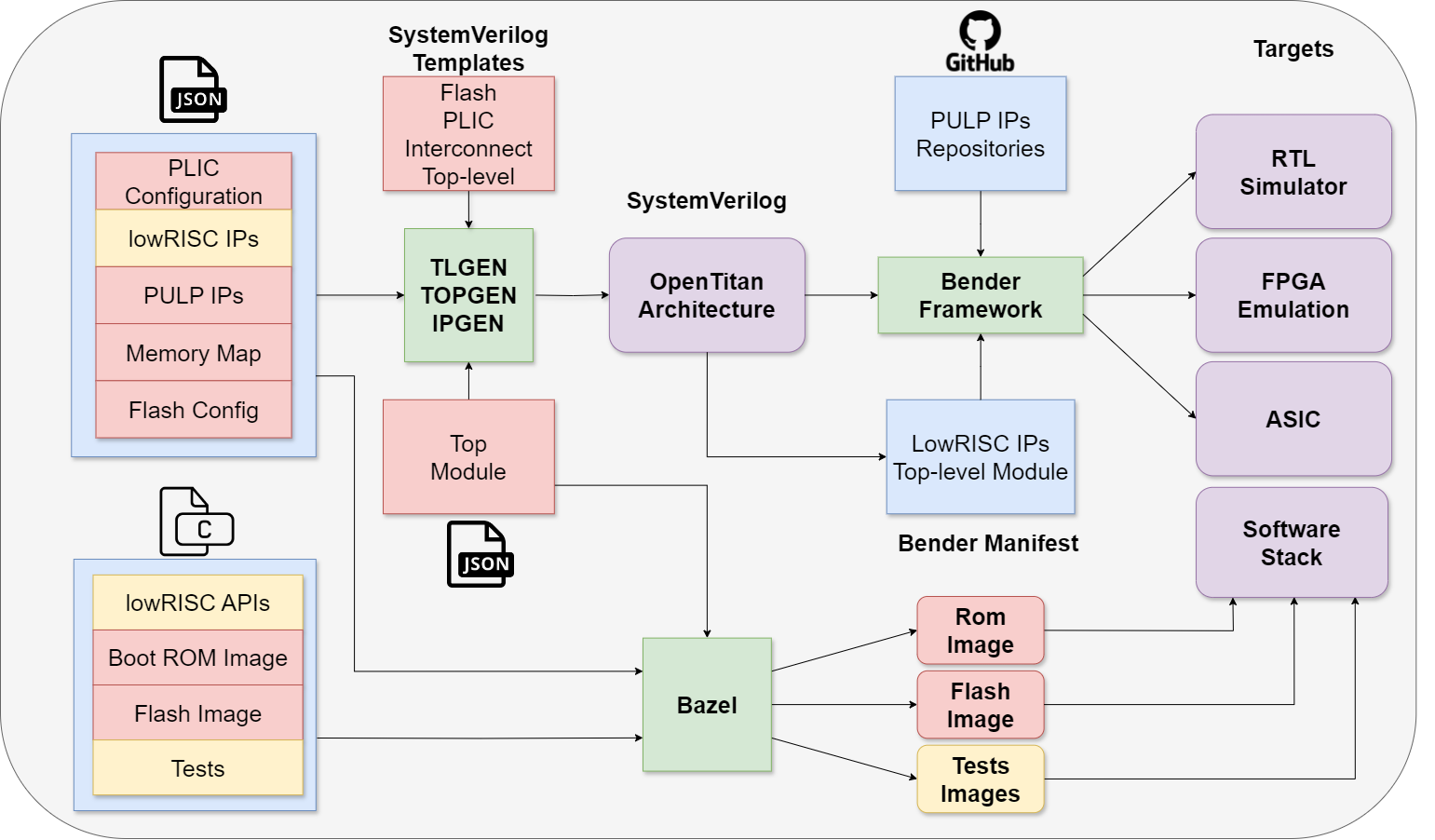}
  \caption{This block diagram shows the methodology with which we deployed OpenTitan in our environment. Red boxes represent the source files we adapted/integrated into our implementation; green boxes represent the stages of the framework; purple boxes represent the outputs of the stages.}
  \label{fig:general-methodology}
\end{figure}

We extended the OpenTitan build and deployment framework, as shown in Figure~\ref{fig:general-methodology}, with PULP utilities to deploy OpenTitan within the environment of the reference system architecture.
We first extended and generated the Earl Grey top level module by exploiting OpenTitan's Python scripts.
We modified the HJSON configuration files for the various IPs and the top-level module and introduced new HJSON files for the PULP IPs we integrated.
Once the top level module is generated with the extensions, we exploit the PULP framework based on Bender to deploy the platform.
We created the Bender manifest, including all the lowRISC IPs' source files and Git external dependencies required for the extensions.
We defined the various implementation targets, allowing us to fetch the OpenTitan repository as a Git dependency into the reference SoC for the integration.
With Bender, it is possible to generate different TCL scripts for different deployment scenarios, encompassing RTL simulations, FPGA emulation, and ASIC.
To compile the boot ROM and Flash images, which are strictly coupled with the APIs provided by lowRISC, we exploited OpenTitan's Bazel-based framework.
The software stack is linked to the RTL implementation, as some definitions, such as flash memory size and the memory map, depend on the HJSON files of the IPs and the top level.
Furthermore, we customized the ROM and flash images as we extended the secure boot framework to the application domain.

\subsection{Extensions for System-level Integration}

\begin{figure}[b!]
  \centering\includegraphics[width=0.7\columnwidth]{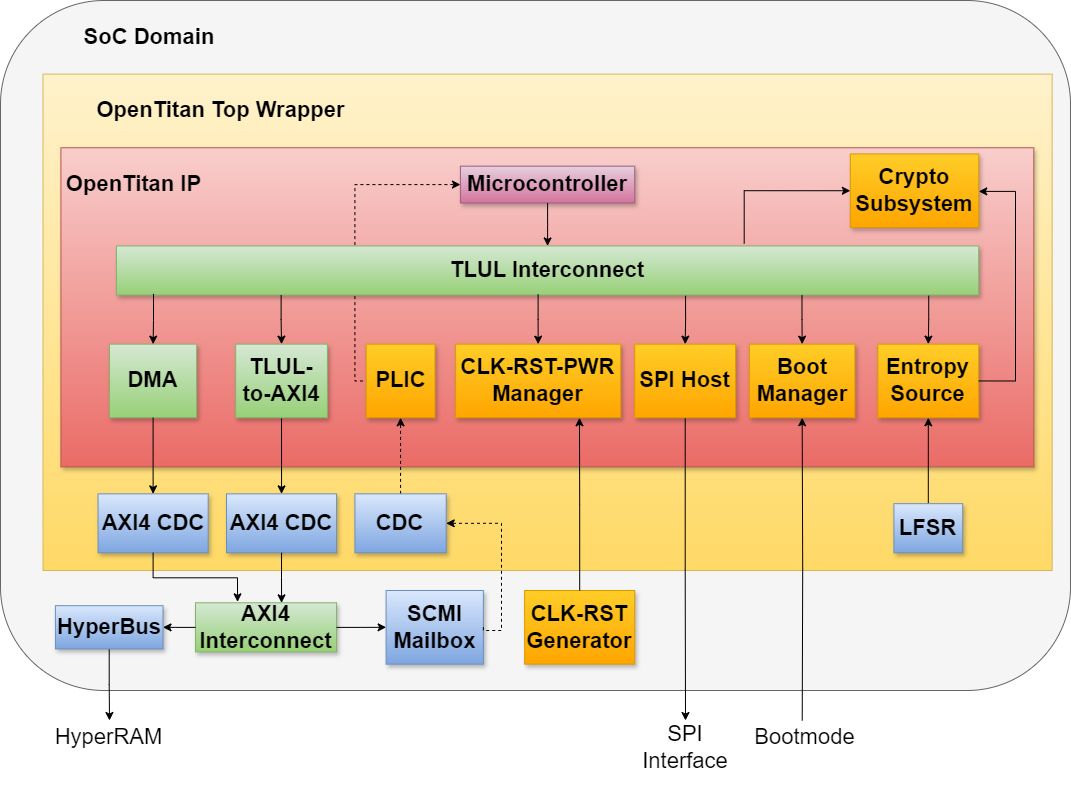}
  \caption{This picture shows the block diagram of the Earl Grey wrapper module.}
  \label{fig:wrapper}
\end{figure}

The focus of our study is the central digital architecture of the Earl Grey OpenTitan implementation, the \texttt{top\_earlgrey} module, thus we did not include in our design the Analog Sensor Top (AST), despite its critical role in OpenTitan's security framework.
Earl Grey is designed to interface with the external environment via standard off-chip communication peripherals, like the SPI interface, being OpenTitan a stand-alone system-on-chip.
While optimal for wired inter-chip communication, these protocols are not inherently designed for intra-SoC messaging in an embedded use case. 
OpenTitan's capacity to implement security services, such as Trusted Execution Environment primitives, necessitates comprehensive access to the host's memory map for real-time integrity assessments and eventually proactively managing the system's behavior in response to detected security breaches.

\begin{figure}[t!]
  \centering\includegraphics[width=\columnwidth]{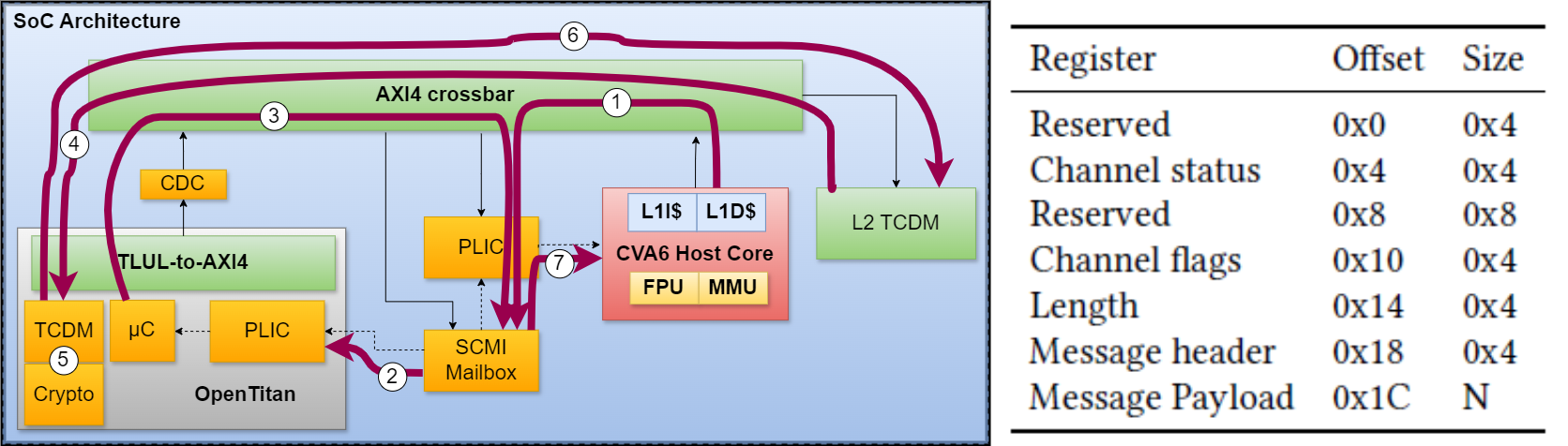}
  \caption{On the left side it is explained the delegation process where the host populates mailbox's registers (1) with data buffer's location, its size, and a distinctive identifier that outlines the required operation. Then, the host raises the mailbox interrupt towards OpenTitan (2). OpenTitan fetches and decodes the command within the mailbox (3) and load the eventual input payload from the external memory (4). At this point, OpenTitan executes the task (5) and stores eventual outputs (6) back in memory before raising the interrupt to the application processor (7). On the right side ,the memory map of the mailbox is shown.}
  \label{fig:mbox_comm}
\end{figure}
We have expanded Earl Grey architecture's connectivity by introducing a new master port into its central interconnect, which merges into the SoC's primary interconnect, thus granting OpenTitan complete visibility over the entire memory map of the SoC. 
To safeguard the host core from tampering with the secure subsystems' content, OpenTitan is integrated with the broader SoC architecture deliberately omitting slave port access.
Notably, Earl Grey's use of the TLUL protocol contrasts with our target architecture adoption of the AXI4\footnote{\url{https://developer.arm.com/documentation/ihi0022/latest/}} standard, especially given TLUL's limitations in high-performance contexts due to its design leaning towards simplicity and straightforward integration.
Such simplicity results in the absence of advanced capabilities such as burst transfers and out-of-order transactions found in AXI4, which enhance data throughput and latency in heterogeneous systems. 
We specifically designed a TLUL-to-AXI4 bridging module, as AXI4 is commonly employed as a standard in heterogeneous architecture like the one we target for OpenTitan integration. 
Typically, the memory map of the SoC is a specification and cannot be changed. 
In this case, one must adapt the OpenTitan default memory map to comply with it.

In scenarios where Earl Grey is integrated into a larger SoC, a wrapper module facilitates its integration by managing the interface between Earl Grey's architecture and the application domain, ensuring that only relevant signals are exposed for system-level integration and handling the interface towards the AST not included in our implementation.
We designed a SystemVerilog wrapper for the Earl Grey architecture to seamlessly integrate it into broader SoCs.
The wrapper module, shown in Figure~\ref{fig:wrapper}, includes a Random Number Generator (RNG) based on Linear Feedback Shift Registers (LFSR), providing the entropy required by OpenTitan.
It also includes Clock Domain Crossing (CDC) stages to synchronize the various interfaces towards the host domain, as the clock domains are different, allowing OpenTitan for a stand-alone physical implementation.
Furthermore, the wrapper is in charge of setting the AST-related inputs to a specific configuration that allows the \texttt{top\_earlgrey} module to bypass the interactions with the AST.

To allow the application domain to request specific services from the Root-of-Trust, such as TEE primitives or cryptographic tasks, we implemented a secure communication channel from the host to Earl Grey without requiring to expose a slave port to be driven by the host.
The communication channel is based on a shared mailbox we designed.
For the mailbox, we opted to comply with a standard commonly employed in heterogeneous architectures, that is ARM System Control and Management Interface (SCMI)~\footnote{\url{https://developer.arm.com/documentation/den0056/d/?lang=en}}.
SCMI is a standardized interface that facilitates communication between system control processors, such as microcontrollers managing system power states and application processors within a SoC.

The mailbox consists of a register file, shown in~\ref{fig:mbox_comm}, on the right side, that complies with the SCMI specification, including two dedicated registers for interrupt-based communication. 
Interactions between the primary processor and the secure module are confined to message exchanges through an interrupt-driven mailbox system.
According to SCMI standards, the mailbox provides two interrupt lines connected to the PLICs of the application processor and OpenTitan, represented by the dotted lines in Figure ~\ref{fig:mbox_comm}.
The former provides the mailbox with a command with pointers to the payloads and destination address, following SCMI standards. 
Then, it raises the interrupt towards OpenTitan.
Received the interrupt from the mailbox, OpenTitan reads and encodes the command, then retrieves the payload from the specified address and processes the task.
Upon completion, OpenTitan returns the results to the specified address and raises back the interrupt towards the application domain.
Figure~\ref{fig:mbox_comm} focuses on OpenTitan's integration within the SoC and how the communication via mailbox happens.
Since the application processor must also access the mailbox, it cannot be instantiated within OpenTitan perimeter.
Conversely, the communication channel would require the host to access the OpenTitan silicon region.
Instead, the mailbox is integrated into the SoC domain and can be accessed by both systems, as OpenTitan has access to the SoC memory map through the TLUL-to-AXI4 bridge we designed. 
OpenTitan's PLIC has been extended to accommodate the new interrupt line from the SCMI mailbox.
The external interrupt is the only way the host has to directly interact with OpenTitan, whose PLIC default configuration ensures that such interrupts are masked and not propagated to OpenTitan's microcontroller, unless specifically configured.
%The external interrupt is the only way the host has to directly interact with OpenTitan, whose PLIC configuration ensures that such interrupts are blocked by default.

\subsection{Optimizing the Cryptographic Engines' Datapath}
\label{subsec:Optimizing the Cryptographic Engines' Datapath}

\begin{figure}[b!]
  \centering\includegraphics[width=0.85\columnwidth]{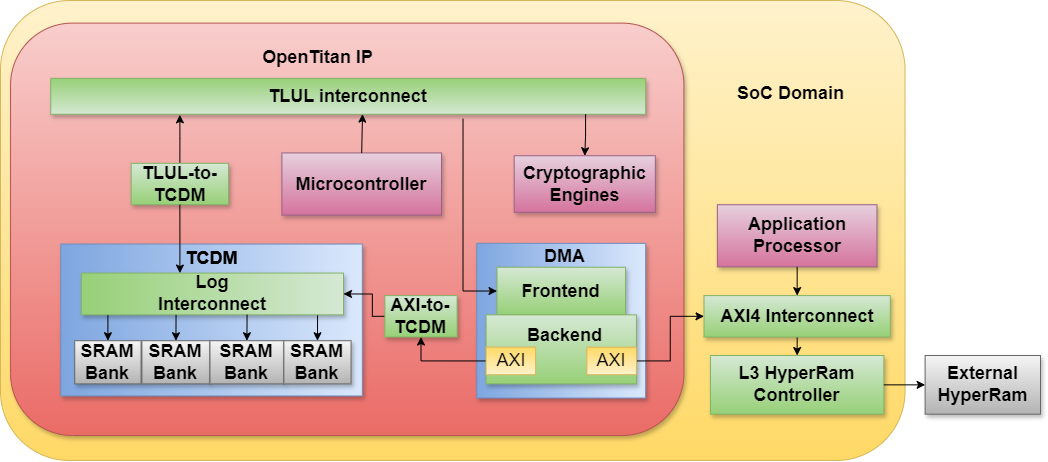}
  \caption{This block diagram shows how DMA and TCDM are integrated in detail.}
  \label{fig:opentitan_ext}
\end{figure}

Earl Grey primarily focuses on security and trustworthiness rather than performance metrics.
The datapath between the accelerators and the memory is inefficient as the interconnect introduces six clock cycles of latency at each transaction.
Moreover, the tiny Ibex core moves data across the OpenTitan platform with loads and stores. 
There is no data movement engine, such as a DMA engine, which can manage data movement automatically, requiring only some configuration from the microcontroller.
This results in sub-optimal utilization of the accelerators, which spend most of the clock cycles waiting for the Ibex core to transport data across the accelerators.
The situation worsens if we consider an external system with a particular memory hierarchy and eventually some protocol conversion and clock domain crossing stages in between.
Accessing a payload in the host memory hierarchy implies a more significant delay, depending on the architecture into which we integrate OpenTitan. 
Since the main bottleneck is the TLUL interconnect, we optimized it to reduce the latency it introduces.
Each FIFO within the interconnect introduces at least one clock cycle of latency, even if the FIFO is empty.
Between each master and each slave, there are at least two FIFOs at the request and response interface, introducing at least four latency clock cycles per transaction.
It is possible to reconfigure and customize the interconnect, imposing the FIFOs to the pass-through mode when empty.
This optimization lands to an average latency in accessing memory of about two clock cycles, enough to saturate the accelerators' bandwidth if the payload is larger than 1KB.

Nevertheless, this does not hold for load and store in memory regions external to OpenTitan IP, as this depends on the specific target architecture, its memory hierarchy, and the datapath. 
Employing the microcontroller to retrieve data from the external memory hierarchy introduces additional clock cycles of latency to the ones already introduced by the TLUL interconnect, wrecking the performance: in total, to handle data transport exploiting TLUL-to-AXI4 bridge, we would require more than 20 clock cycles on average, introduced by three CDC stages, bridging module between AXI4 and TLUL, and eventual data dependencies stalling the microcontroller's pipeline.
To approach the performance observed  when the payload is confined within OpenTitan, we extended Earl Grey architecture to accommodate
an open-source DMA\cite{idma} engine, dedicated exclusively to handling data movement across the OpenTitan perimeter, from and towards the external memory hierarchy. 
Additionally to the DMA, we integrated 2KB of TCDM dedicated exclusively to the accelerators, where 8x 4KB memory banks are connected to a logarithmic interconnect accommodating two master ports, one for the microcontroller and the other for the DMA. 
This TCDM provides a memory region with high bandwidth access, thanks to the logarithmic interconnect. 
We opted for 32KB of TCDM because it is large enough for the experiments we carried on, but it can be adjusted at compile time, depending on the available silicon area. Furthermore, the data within OpenTitan perimeter is moved throughout loads and stores performed by Ibex microcontroller on the TLUL interconnect, which does not support bursts transactions, while the DMA does. The performance is strictly bounded by Ibex moving data. On the other hand, the DMA can exploit TCDM’s log interconnect bandwidth, but only to fetch/store back data from/to external memory, not during the cryptographic operations, where data is moved between the TCDM and the accelerators by Ibex. Furthermore, again the bandwidth is bounded by the latency on reading/writing from/to the L3 memory. It has not been included any scramble mechanism and its corresponding logic for the TCDM, as the TCDM IP we have available does not support such mechanisms yet. The TLUL interconnect directly drives, through a protocol converter, the TCDM's master port. Ibex microcontroller accesses the TCDM with the same latency observed in other embedded memory regions, while the DMA can exploit TCDM’s full bandwidth, but still it cannot saturate it because of the latency on accessing the L3.
Ibex microcontroller accesses the TCDM with the same latency observed in other embedded memory regions, while the DMA can exploit TCDM’s full bandwidth.
For this reason, once the DMA has moved the payload into the TCDM, the memory accesses by the Ibex introduces 2 clock cycles of
latency instead of more than 20, at the cost of the overhead of waiting the DMA to complete the transaction. 
This overhead is strictly dependent on the payload size, and can be hidden within the task execution, implementing double buffering techniques. 
In the block diagram shown in Figure \ref{fig:opentitan_ext}, the integration of the two new modules are described.
The DMA we integrated consists of two main blocks, which are frontend and backend.
The frontend consists of a register file from which the microcontroller can configure the source and destination address, set the payload length, and other configurations. 
The backend is driven by the frontend and exposes two AXI4 master port and manages the data movement from source to destination.
One of the ports is connected to the external AXI4 interconnect, while the second one is connected to the high bandwidth TCDM, through a dedicated protocol converter. 
Exploiting the DMA along with the interconnect’s optimizations, the
performance enhances by a factor of 2.7$x$ and 1.6$x$ for HMAC/SHA-256 and AES respectively.
\section{OpenTitan as a Silicon-ready IP}

This section outlines the modifications made to the Earl Grey architecture to facilitate its deployment.
We discuss the reasons behind the forfeiture of the AST subsystem, and we address cases in which specific IPs in the target technology node, like the flash IP and the eFuse, are not available or when facing limitations due to the constraints in the silicon area.
Moreover, we discuss why we opted to replace the eFuse memory with a mask-ROM.
Most importantly, we discuss how we modified the secure boot framework to adapt to the abovementioned limitations and extended the secure boot to the reference architecture. We also introduced a debug boot mode to provide a straightforward framework for conducting research on the OpenTitan platform.

\subsection{OpenTitan as a Silicon-ready Prototyping Vehicle}

OpenTitan is designed to be a product that provides a tamper detection mechanism based on analog sensors handled by the AST. 
The AST's role is to enhance the platform's security and monitoring capabilities by providing direct and accurate measurements of environmental and system parameters critical for maintaining operational integrity.
For a research platform focused on the digital aspect of the architecture from a system-level viewpoint, such a feature can be forfeited to simplify the implementation of the digital platform.
Moreover, it could be challenging for research centers with limited resources to implement all the features provided by the OpenTitan project.
It is possible to conduct research on the digital system architecture without employing the AST, despite it being a fundamental building block for the OpenTitan security model.
The same considerations above hold also for the entropy source and the PUF,  which is required to derive a device unique secret (i.e. the Silicon Creator Root Key) using an underlying entropy source (i.e. ring oscillators).

\subsection{Lifecycle State's Customization}

Deploying OpenTitan requires careful attention to the lifecycle state, stored within the OTP memory, which relies on eFuses.
Initially, the OTP remains blank at tape-out, requiring the chip's bring-up phase to involve the lifecycle manager's dedicated JTAG interface.
The microcontroller's fetch is enabled within this lifecycle state, and the main JTAG interface is disabled.
Users input specific tokens through the lifecycle manager JTAG interface, transitioning the system from the "Scratch" state to the "Test-Unlocked" state, where core execution and main JTAG functionalities are enabled.
The "Test-Unlocked" state is suitable for testing but not for deployment in a real scenario, as the JTAG, for instance, is always enabled.
In "Test-Unlocked" state, the lifecycle manager activates the microcontroller's fetch-enable, the main JTAG, and other features previously inactive in "Scratch" state. 
However, navigating through this phase is challenging during bring-up and was not deemed essential for harnessing primary security of the central digital system of OpenTitan.
Furthermore, not every research institute can access non-standard IPs such as eFuse, PUFs or non-volatile memories for specific technology nodes.
Since the system-on-chip object of this work is not a product, but a prototype for academic research, we opted to preset the lifecycle state to "Test-Unlocked" and preload cryptographic seeds and keys, substituting the OTP eFuses with a mask-ROM reflecting the post-bring-up state as per the OpenTitan guidelines. 
This approach significantly streamlines the system's bring-up process, allowing it to load and test arbitrary software through the JTAG interface directly without going through the lifecycle state framework from scratch.
The cost for this simplification is the forfeiture of certain functionalities, such as silicon ownership transfer and lifecycle transactions, because eFuses are replaced with an immutable ROM.
While these functionalities are essential for a product, they are not essential for a research platform.

\subsection{OpenTitan Flash Controller Extension to Support an External Flash}

\begin{figure}[t!]
  \centering\includegraphics[width=0.8\columnwidth]{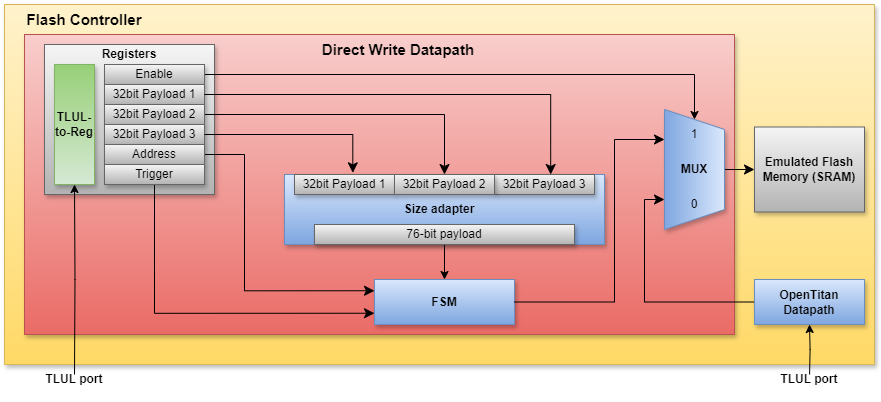}
  \caption{This block diagram shows the datapath extension we implemented to the flash controller.}
  \label{fig:datapath_flash}
\end{figure}

The OpenTitan design embeds a flash memory to store the ROM extension boot code, the second boot layer after the secure boot routine. This allows for updates to this ROM extension partition after the tape-out. However, there is no possibility to populate the embedded Flash with the content of an external SPI Flash before running the secure boot routine. Moreover, the embedded non-standard IPs, such as the flash IP, in the target technology node might be unavailable, creating issues for research projects. Moreover, the OpenTitan default configuration is provided with 2MB of flash memory, which could be challenging to fit within the available silicon area, a specification of the reference SoC.
We present an extension to the flash controller datapath, described in Figure~\ref{fig:datapath_flash}, which does not require the presence of an embedded flash IP and allows loading code from an external SPI flash.
We also reduced the data partition's bank size from 2MB to 64KB to fit within the silicon area imposed by the specifications.

This approach involves replacing the flash banks with SRAM banks, emulating the flash memory by loading to the SRAMs the flash image before each boot attempt.
The datapath of the flash controller is updated with the new memory size, exploiting OpenTitan's scripts.
This bypass mode does not impact the secure boot framework.
The flash controller does not allow direct write, which the controller's FSM must handle. 
The microcontroller must program the flash controller's register file to issue a write on the flash memory.
Since we emulate the Flash with SRAM banks, it is possible to perform direct writes.
This requires bypassing the direct write protections of the flash controller, allowing direct writes to the SRAM banks, which emulate the flash memory.
We extended the flash controller, introducing an alternative datapath, multiplexed with the main one to allow direct writes.
The flash controller introduces 6 bits of ECC and integrity bits to each 32-bit word, extending the word length to 38 bits.
It concatenates two 38-bit words into a unique 76-bit word, which must be the data width of the SRAM banks emulating the Flash.
A datapath to handle this different data width presents a significant challenge since the TLUL interface is designed for 32-bit transactions.
This datapath consists of a memory-mapped register file and an FSM driving the SRAM's interface.
The register file includes three registers to accommodate the 76-bit word, one address register pointing to a specific SRAM entry, and a trigger register to issue the FSM after the registers are populated.
Furthermore, there is a register driving the multiplexer between the two datapaths.
Before the secure boot starts, the microcontroller drives the SPI interface to fetch the encrypted flash code from the external Flash and copy it into the main SRAM.
Then, it exploits the alternative datapath to move the image from the main SRAM toward the emulated Flash.
The microcontroller splits the 76-bit payload into three write transactions to the three payload registers.
The last register is written with the last 12 bits of the payload, while the remaining 20 bits are padded to zero as they are unused.
The microcontroller then writes the address register with the specific memory cell and raises the trigger signal, activating the FSM.
Upon activation, the datapath reads and concatenates the payload registers, excluding the extra 20 bits.
It then performs one unique 76-bit write transaction to the interface of the emulated Flash at the address provided by the register.
This is repeated for every 76-bit word read from the external Flash.
Correspondingly, the boot ROM code has been modified to accommodate this process.

\begin{figure}[t!]
  \centering\includegraphics[width=\columnwidth]{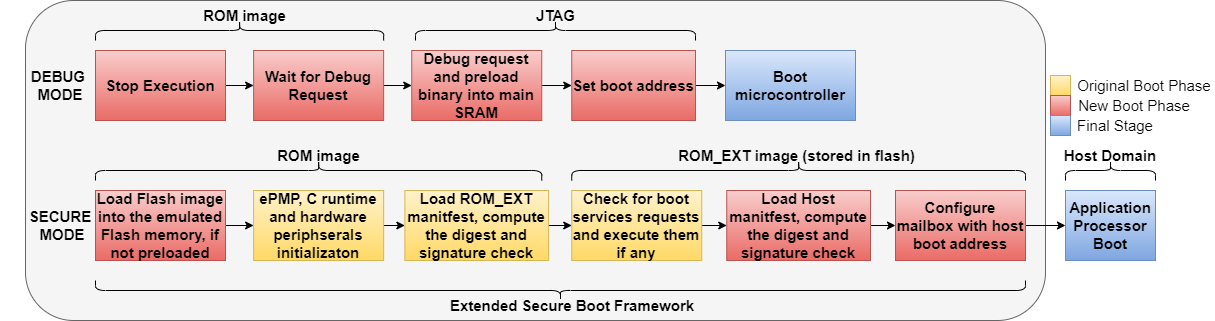}
  \caption{This block diagram describes the two boot modes we introduced into OpenTitan.}
  \label{fig:bootmodes}
\end{figure}

\subsection{OpenTitan's Bootflow Customization}

The OpenTitan framework is designed without the flexibility to select boot modes. 
It begins executing the boot ROM image immediately upon receiving the fetch-enable signal from the power manager's FSM, on top of successful integrity verification of the OTP's and ROM's images.
The ROM software stack is purposely built to initiate the secure boot process directly, lacking any boot mode selection feature.
Consequently, we introduced a new boot mode, an alternative to the secure one implemented by default.
A boot manager module was integrated into the OpenTitan architecture and its memory map to accommodate these two modes alongside a new input signal that dictates the boot mode. 
This module includes a register file with one register specifically for capturing an external boot-select signal. 
We revised the ROM code to read the boot mode register before proceeding with the corresponding boot mode.
Integrating the boot manager into the system was seamless, only necessitating an additional port on the TLUL interconnect for the boot manager.
Furthermore, we updated the boot ROM code for fetching content from an external SPI flash to store in the emulated Flash before commencing with the secure boot process, as we extended the flash controller datapath and replaced the flash IP with SRAM.
The ROM waits for debug commands in debug Mode through the JTAG interface.
Via JTAG, it is possible to load code in OpenTitan's main memory and then program the microcontroller to jump to it.
On the other hand, secure mode commences the standard ROM execution, starting with the secure boot.
Debug Mode provides a simplified setting for IP testing, enabling the direct loading and execution of arbitrary code in the main SRAM, thereby bypassing secure boot constraints.
Figure~\ref{fig:bootmodes} shows how the two boot modes are implemented, highlighting each step, which is the source of the code being executed by the microcontroller.

\begin{figure}[t!]
  \centering\includegraphics[width=\columnwidth]{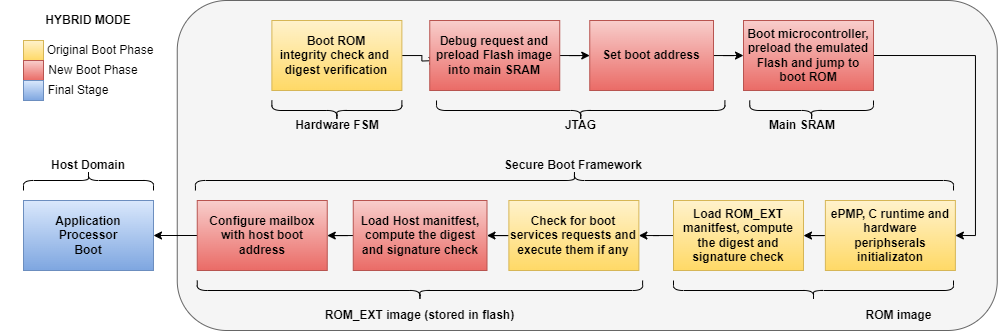}
  \caption{This block diagram shows the Hybrid boot mode, executing the secure boot in Debug Mode by preloading OpenTitan through the JTAG instead of retrieving the flash image from the SPI interface.}
  \label{fig:hybrid-boot}
\end{figure}

Nevertheless, the microcontroller can preload the emulated Flash even in Debug mode. 
This possibility enables a third hybrid boot mode, which allows the execution of the secure boot in debug mode without needing an external SPI flash.
In debug mode, it is possible to program the microcontroller to load the flash image, preloaded via JTAG, from the main SRAM and copy it to the emulated Flash, driving the alternative datapath's register file.
At this point, the microcontroller jumps directly to the boot ROM address as soon as the emulated Flash is preloaded with its image.
The secure boot from this point follows the same steps as in secure mode.
The only difference is that the microcontroller may bypass reading the flash image via SPI, as the image is already loaded.
To tell the boot code that it shall not load the flash image from the SPI interface, we introduced a further software-writable register to the boot manager for secure boot.
The microcontroller writes this new register after moving the image from the main SRAM, preloaded via JTAG, to the emulated Flash.
After checking the boot mode from the external pad, the boot ROM code then checks this register at the very first instructions.
Executing the secure boot in this hybrid boot mode is the most straightforward approach for verifying the functionality of the secure boot, as it only requires the JTAG connection towards OpenTitan.
Figure~\ref{fig:hybrid-boot} presents a block diagram of the hybrid boot mode, similar to the one in Figure~\ref{fig:bootmodes}.

The secure boot is split into two phases: a first stage where the microcontroller executes the image stored in the boot ROM and a second stage where it runs the stored in the emulated Flash, which can accommodate an extension of the ROM image and different partitions for different boot stages.
Each stage is responsible for validating the subsequent boot stage and configuring the PMPs accordingly.
The ROM stage handles the cryptographic measures and verification on the flash boot stage.
The only mandatory boot stages are the boot ROM and its extension stored in the emulated Flash.

After these steps, the Flash image can either verify and boot on the OpenTitan platform's different boot stages, such as boot loaders, or directly verify the host domain firmware and boot the application processor.
We extended the OpenTitan boot flow to the application processor, exploiting the mailbox and its interrupt, as shown in Figure~\ref{fig:bootmodes}.
Having access to the SoC memory map, OpenTitan can configure the PLIC of the application processor, enabling the mailbox to interrupt and perform further configurations.
The boot ROM of the application processor is programmed to wait for the mailbox interrupt or a debug request via JTAG.
OpenTitan populates its payload register with the boot address before raising the interrupt of the mailbox. 
Once the mailbox interrupt is received, the application processor reads the boot address from the mailbox and jumps to it.
This boot flow extends OpenTitan's secure boot to the application domain, ensuring that the firmware the host executes is authenticated on top of the authentication of the lower boot stages within the OpenTitan Root-of-Trust.
\section{Results}

This section discusses the results of our study. 
We show the physical implementation of OpenTitan, the system architecture in a specific technology node, and the distribution of area occupation of the OpenTitan's submodules and the SoC.
We assessed and compared the performance of OpenTitan as a cryptographic engine both on the base architecture and on the extended one with the datapath described in Section~\ref{subsec:Optimizing the Cryptographic Engines' Datapath}, showing that our extensions improve the performance of the baseline system, saturating the accelerators' bandwidth.

\subsection{Physical Implementation}
%to finalize integration of this image and its references in the text
\begin{figure}[]
  \centering\includegraphics[width=\columnwidth]{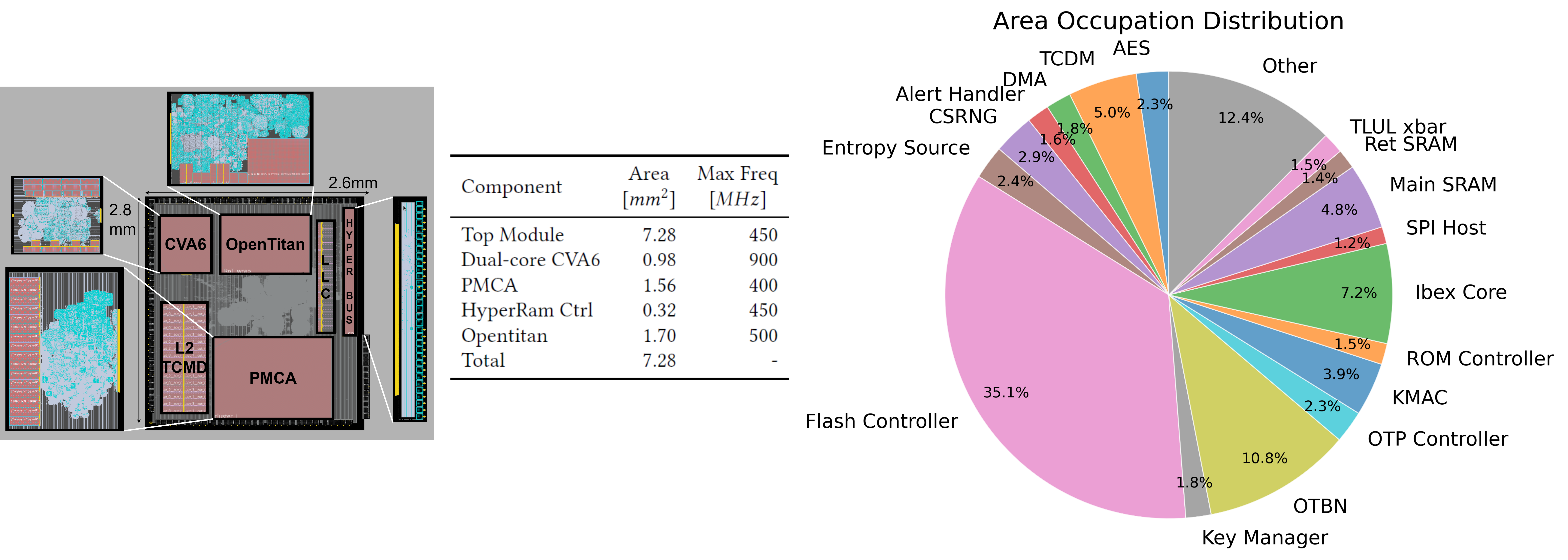}
  \caption{On the left side, the floorplan of the SoC. The area occupation and target frequencies of the various SoC subsystems are in the middle. The area occupation's distribution over the various OpenTitan components is on the right.}
  \label{fig:area-occupation}
\end{figure}

We implemented the SoC in Global Foundries 22nm FD-SOI technology node.
The various subsystems composing the SoC, including OpenTitan IP, are implemented as a hardened macro, separating their clock domains from the host one and allowing for a parallel stand-alone implementation process for all the subsystems.
The synthesis of the design was performed using the Synopsys Design Compiler 2020.09, while the back-end design flow was executed with Cadence Innovus 20.12.000.
We signed off the design considering the worst-case operating corner for a nominal supply voltage of 0.8 V (SS, 0.72 V, 125°C/-40°C), targeting 500MHz for OpenTitan's clock domain and 900MHz for the application processor, as shown in the table of Figure~\ref{fig:area-occupation}.
Figure~\ref{fig:area-occupation}, on the right side, shows a pie chart describing how the area occupation is distributed across OpenTitan's submodules.
The plot explicitly shows the submodules that occupy more than 1\% of the area.
Moreover, on the left, the floorplan of the SoC is shown. 
Despite reducing the emulated flash memory's size to 64KB, the flash controller occupies most of the silicon area, about 35\%.
The flash controller has two 32KB data banks to allow for firmware updates with A-B partitioning.
Three information partition memory banks are employed for each data partition, which we did not reduce as this is more complex than reducing the data partitions.
The largest cryptographic accelerator is the OTBN, which occupies about 11\% of the total OpenTitan area. In contrast, the other accelerators, such as KMAC, AES, and HMAC, occupy 3.9\% of the area in the worst case.
The plot shows that the DMA and TCDM introduce less than 5\% % of area overhead.
The table in Figure~\ref{fig:area-occupation} shows the area occupation and the maximum frequency of the various domains of the SoC, including OpenTitan.
The total silicon area available for the design is 2.8$mm^2$ by 2.6$mm^2$, resulting in 7.28$mm^2$.
While its area has been reduced as much as possible, OpenTitan occupies the largest portion of the design, making its integration in low-area design unfeasible.
Further customization would be required to reduce the area occupation of OpenTitan, namely further reducing the flash memory size, acting on information partition's banks, and removing the OTBN co-processor.
The latter can be removed without impacting secure boot and security features, as lowRISC provided a framework to run the RSA algorithm in software on the Ibex core at the cost of additional latency.

\subsection{Benchmarking Methodology}

We extended the benchmarks Parisi et al. utilized for \cite{parisi2024assessing} to assess the performance of OpenTitan integrated in a broader system, when used as a security workload accelerator taking into account the memory hierarchy of the platform into which OpenTitan is integrated and to assess the speed-up derived by our extensions, comparing it against the performance of executing similar workloads on the application processor without any hardware accelerator.
The benchmarks are run on two versions of OpenTitan: the base architecture, including only the TLUL-to-AXI4 bridge as extension to access external memory, and the enhanced architecture encompassing the DMA, TCDM, and optimizations to the TLUL interconnect.
We explored a range of security functions critical for securing communication channels and safeguarding data.
For each identified function, we developed two implementations: one intended for software execution on the CVA6 and another for execution on the OpenTitan subsystem, designed to exploit OpenTitan hardware acceleration features.
The security functions evaluated in this study are SHA-256, HMAC, and AES-256 (in CBC mode).
The software benchmarks were built upon BearSSL, an open-source library that implements cryptographic algorithms essential for a complete SSL/TLS communication protocol. 
We utilized libraries tailored for performance, rather than security as the latter would be slower. For instance, the AES implementation is not constant-time and unprotected. For this reason, the performance numbers on the software implementation represent an upper bound. The software implementations have been executed only once on the L3 memory
All implemented cryptographic algorithms are written in pure C without using ISA extensions.
In assessing OpenTitan's hardware accelerators, we designed benchmarks for two distinct scenarios: one where input data is stored in external L3 memory and another involving data stored within OpenTitan's private L1 memory. We did choose to carry these experiments only involving the L3 memory and not the L2 memory as the former is more representative of a real use-case, the L3 is the main memory of the system. The assumption is that in all the experiments, OpenTitan accesses the L3 with a hit-rate of always 100\% on the LLC. We define the performance metric to be the number of clock cycles required to process one byte of payload ($Clks/B$).
The first scenario is typical of applications where the host core utilizes OpenTitan as a cryptographic accelerator, involving the host core providing OpenTitan with data pointers and metadata. 
The second scenario addresses operations on sensitive data already transferred to OpenTitan's private scratchpad, highlighting how the memory hierarchy impacts performance, with operations on the Root-of-Trust's private scratchpad being faster than those on main memory external to OpenTitan. 
In both cases, cryptographic operation results are returned to the same memory hierarchy level from which the input data was sourced without exploring cross-hierarchy data movements.
The benchmarks are implemented in C, directly configuring the accelerators without exploiting the test functions included in the official OpenTitan software suite.
The latter, while illustrative and fundamental for system verification, are not optimized for performance as they perform unnecessary memory accesses that could degrade the maximum performance of the hardware accelerators. 
Moreover, we carefully managed scenarios requiring the software to wait for accelerator readiness (e.g., full input FIFOs or need for reconfiguration) through polling to eliminate delays caused by interrupt handling inefficiencies and wake-up times.
We then compare the results of the benchmarks on the two versions of the architecture to highlight the bottlenecks in the base architecture and how we tackle them.
Moreover, we compare the difference in performance obtained when retrieving the payload from the embedded memory or the external one, highlighting the impact of the memory hierarchy on the performance.
\begin{table*}
    \centering
    \caption{Meaning of the semantic labels used to annotate the benchmark execution traces.}
    \resizebox{\textwidth}{!}{
    \label{table:labelling}
    \begin{tabular}{clll}

        \toprule

        Type & IP & Label & Description \\

        \midrule

        \multirow{9}{*}{\rotatebox[origin=c]{90}{Semantic}}
        & \multirow{4}{*}{HMAC} & Configure & Configure the accelerator in SHA-256 or HMAC mode and load the secret key if necessary. \\
        &                       & Digest & Load data from memory and push it into the accelerator FIFO making it ready for processing. \\
        &                       & Wait   & Wait for the accelerator to be ready to accept further data. \\
        &                       & Finalize  & Pad the message, finalize digest computation and copy the result back to memory. \\
        \cmidrule(lr{0.5em}){2-4}
        & \multirow{3}{*}{AES}  & Configure & Initialize the accelerator, set the mode of operations, the key, and the initialization vector. \\
        &                       & Cipher & Push the next data block into the the accelerator and writes the previous block back to memory. \\
        &                       & Wait   & Wait for the accelerator to be ready to accept further data. \\
        \cmidrule(lr{0.5em}){2-4}
        & \multirow{2}{*}{ALL} & DMA   & Issue a DMA transaction. \\
        &                      & DMA Wait   & Wait for DMA to complete the transaction. \\

        \bottomrule

    \end{tabular}
    }
\end{table*}
The analysis was conducted using Mentor QuestaSim.
Despite the slower operation of simulation compared to FPGA emulation, which is limited in its ability to record cycle-accurate execution traces for extensive clock cycles, RTL simulations offer the capability to capture assembly execution traces for the entirety of the simulation.
The performance evaluation of the benchmark executed on OpenTitan involves a detailed analysis that combines data from execution traces generated by the RTL simulator and an annotated disassembly of the benchmark software. 
Each assembly instruction in this annotated version is assigned a semantic label, as outlined in Table~\ref{table:labelling}. 
We extended the benchmarks introducing a new label for the DMA instructions in addition to the ones already exploited in \cite{parisi2024assessing} and introducing the instructions to the benchmarks to issue and wait the DMA when the payload is located outside OpenTitan in L3 memory.
The labels indicate various stages of the benchmarks, including the setup and tear-down of the accelerators, initiating the DMA if employed or awaiting its completion, and pausing the microcontroller for the accelerators to process the payload.
This approach facilitates the identification of stages where execution cycles are predominantly consumed, pinpointing whether a cryptographic workload's performance bottleneck is due to data input latency into the accelerators' FIFOs or the bandwidth limitations of the hardware IP in delivering the results of requested operations.
By integrating this labeled information with the cycle cost associated with executing each instruction as per the execution trace, we gain a comprehensive view of the operations that impose the highest cost when utilizing a particular accelerator. 
This method provides a granular understanding of accelerator-specific performance and offers an indirect evaluation of the overall platform architecture and the efficiency of its memory hierarchy.
The performance metrics are derived from the assembly execution traces generated post-benchmark, detailing the frequency and cycle count per instruction and the total cycles required for the benchmark completion.

\subsection{Benchmarks Results}

\begin{table}[htbp]
    \centering
    \begin{minipage}[t]{.45\linewidth}
        \centering % This centers the content of the minipage
        \caption{Memory access cost statistics in cycles.}
        \label{table:memory_access_cost}
        \begin{tabular}{lrr}
            \toprule
            {Memory} & {Base} & {Extended} \\
            \midrule
            L1       & 6.0    & 2.0       \\
            L3       & 23.0   & 23.0      \\
            TCDM     & {--}   & 2.0       \\
            \addlinespace
                        \midrule
            DMA Latency [$Clks/B$] & \multicolumn{2}{c}{1.4} \\
            \bottomrule
        \end{tabular}
    \end{minipage}%
    \hfill
    \begin{minipage}[t]{.45\linewidth}
        \centering % This centers the content of the minipage
        \captionof{table}{Nominal bandwidth for cryptographic operations and DMA.}
        \label{table:cryptographic_bandwidth}
        \begin{tabular}{rr}
            \toprule
            {Module} & {Bandwidth [$Clks/B$]}\\
            \midrule
            AES       & 4.5 \\ % Replace 'Value1' with the actual value for AES
            HMAC      & 1.25 \\ % Replace 'Value2' with the actual value for HMAC
            DMA       & 0.25 \\
            \bottomrule
        \end{tabular}
    \end{minipage}
\end{table}

Our initial evaluation investigates the highest memory bandwidth the microcontroller can achieve when interfacing with the external L3 memory and the Root-of-Trust private L1 memory.
Given the absence of an on-chip DMA dedicated to transferring data between the various entities within OpenTitan, the microcontroller is tasked with manually managing data transfers from the memory hierarchy to the accelerator FIFO. 
The DMA we integrated is dedicated to transferring data across OpenTitan perimeters from and toward the external memory hierarchy.
Thus, understanding its efficiency in data transfer management is essential, as it could become a bottleneck for accelerator performance in scenarios where memory bandwidth is a limiting factor.
We evaluated the bandwidths of the DMA and the accelerators and compared them with the nominal values obtained by the specifications.
Table~\ref{table:memory_access_cost} presents data on the memory access costs of both the base architecture and of the extended one spent by Ibex microcontroller for each access to either the OpenTitan private scratchpad(L1) and the system RAM (L3), including also the bandwidth we measured for the DMA. 
This data was extracted by aggregating memory access times from various benchmarks, each targeting a different memory type. 
Concerning the DMA bandwidth, we relied on the simulation's waveforms.
Parisi et al. \cite{parisi2024assessing} reveal that accesses to the OpenTitan private scratchpad or any accelerator FIFO average around six cycles, whereas accessing system RAM with a LLC hit takes about 23 cycles when considering the base architecture with the TLUL-to-AXI4 bridge. 
When considering the extended architecture, we observe that the accesses to the embedded SRAM improved by a factor of 3$x$.
Accessing the TCDM requires the same amount of clock cycles as the embedded main SRAM, as both lay into the same hierarchic level and share the same interconnect.
Concerning the DMA, the results show that it requires 1.4 $Clks/B$ to move 1Byte of data. 
Ideally, the DMA should move 4Bytes of data per clock cycle. However, in our integration, we introduced three clock domain crossing stages, introducing about six clock cycles of latency for each 4Bytes transaction, which is six times slower than the nominal conditions.
This result is consistent, as by dividing the DMA bandwidth of 1.4 $Clks/B$ by 6, we obtain 0.23 $Clks/B$, which roughly approximates its nominal bandwidth of 0.25 $Clks/B$.
Table~\ref{table:cryptographic_bandwidth} estimates the minimum cycles per byte that the HMAC and AES accelerators can achieve, including the nominal bandwidth of the DMA. 
The accelerators' estimate is calculated based on the documentation provided by lowRISC, where AES encrypts a 16Byte payload in 77 clock cycles and the HMAC processes 64Bytes in 80 clock cycles.
This results in a minimum bandwidth of 4.5 $Clks/B$ for the AES and 1.25 $Clks/B$ for the HMAC.

\begin{table*}[t]
    \centering
    \caption{Analysis of OpenTitan HMAC and AES accelerator performance on the extended architecture for various payloads}
    \label{tab:opentitan_performance}
    \resizebox{\textwidth}{!}{%
    \begin{tabular}{llrrrr|rrrrr}

        \toprule
        & & \multicolumn{4}{c}{L1 [\% Cycles]} & \multicolumn{4}{c}{L3 [\% Cycles]} \\
        \cmidrule(lr){3-6} \cmidrule(lr){7-10}
        & Operation & 64 Bytes & 256 Bytes & 1024 Bytes & 4096 Bytes & 64 Bytes & 256 Bytes & 1024 Bytes & 4096 Bytes \\
        \midrule

        \multirow{7}{*}{\rotatebox[origin=c]{90}{\textbf{SHA-256}}}
        & Configure               & 2.9  & 1.7  & 0.6  & 0.2  & 2.4 & 1.9 & 1.1 & 0.4 \\
        & Digest               & 40.8 & 63.2 & 82.6 & 90.9 & 7.6 & 14.2 &  26.1 & 34.1 \\
        & Wait                 & 15.8 & 9.3  & 3.5 & 1.0 & 13.5 & 10.6 & 5.9 & 2.0 \\
        & Finalize                & 40.5 & 25.8 & 13.3 & 8.0 & 6.5 & 5.7 & 4.9 & 4.2 \\
        & DMA                  & -    & -    & -   & -  & 0.9 & 0.7 & 0.4 & 0.1 \\
        & DMA Wait             & -    & -    & -   & -  & 69.1 & 66.8 & 61.7 & 59.2 \\
        \midrule
        & \textbf{Performance [$Clks/B$]} & \textbf{5.3} & \textbf{2.3} & \textbf{1.5} & \textbf{1.3}  & \textbf{32.9}&\textbf{10.4} & \textbf{4.7} & \textbf{3.4} \\
        \midrule

        \multirow{7}{*}{\rotatebox[origin=c]{90}{\textbf{HMAC}}}
        & Configure & 7.8 & 5.4 & 2.4 & 0.8 & 8.4 & 5.5 & 2.3 & 0.7 \\
        & Digest & 25.3 & 46.2 & 72.5 & 83.7 & 14.2 & 23.3 & 33.3 & 38.0 \\
        & Wait   & 16.2 & 11.1 & 5.0 & 1.6 & 31.0 & 20.1 & 8.5 & 2.5 \\
        & Final  & 50.6 & 37.3 & 20.0 & 10.3 & 26.8 & 19.3 & 10.5 & 5.8 \\
        & DMA   & - & - & - & - & 1.9 & 1.2 & 0.5 & 0.1 \\
        & DMA Wait & - & - & - & - & 17.7 & 30.6 & 44.9 & 52.9  \\
        \midrule
        & \textbf{Performance [$Clks/B$]}  &\textbf{ 8.6 }& \textbf{3.1} &\textbf{ 1.7} & \textbf{1.4} & \textbf{16.0} & \textbf{6.1} & \textbf{3.7} & \textbf{3.1} \\
        \midrule

        \multirow{6}{*}{\rotatebox[origin=c]{90}{\textbf{AES-256}}}
        & Configure & 27.6 & 10.2 & 2.9 & 0.7 & 20.9 & 6.7 & 1.8 & 0.5 \\
        & Cipher & 32.6 & 45.2 & 50.4 & 52.0 & 23.7 & 28.6 & 30.2 & 31.1 \\
        & Wait   & 39.8 & 44.6 & 46.7 & 47.3 & 28.4 & 28.2 & 28.2 & 27.1 \\
        & DMA   & - & - & - & - & 2.7 & 0.9 & 0.2 & 0.1 \\
        & DMA Wait & - & - & - & - & 24.3 & 35.6 & 39.5 & 41.5 \\
        \midrule
        & \textbf{Performance [$Clks/B$]}    & \textbf{7.9} & \textbf{5.3} & \textbf{4.7} & \textbf{4.6 }& \textbf{10.9} & \textbf{8.5} & \textbf{7.8} & \textbf{7.8} \\

        \bottomrule

    \end{tabular}%
    }
\end{table*}

\begin{figure}[t!]
  \centering\includegraphics[width=\columnwidth]{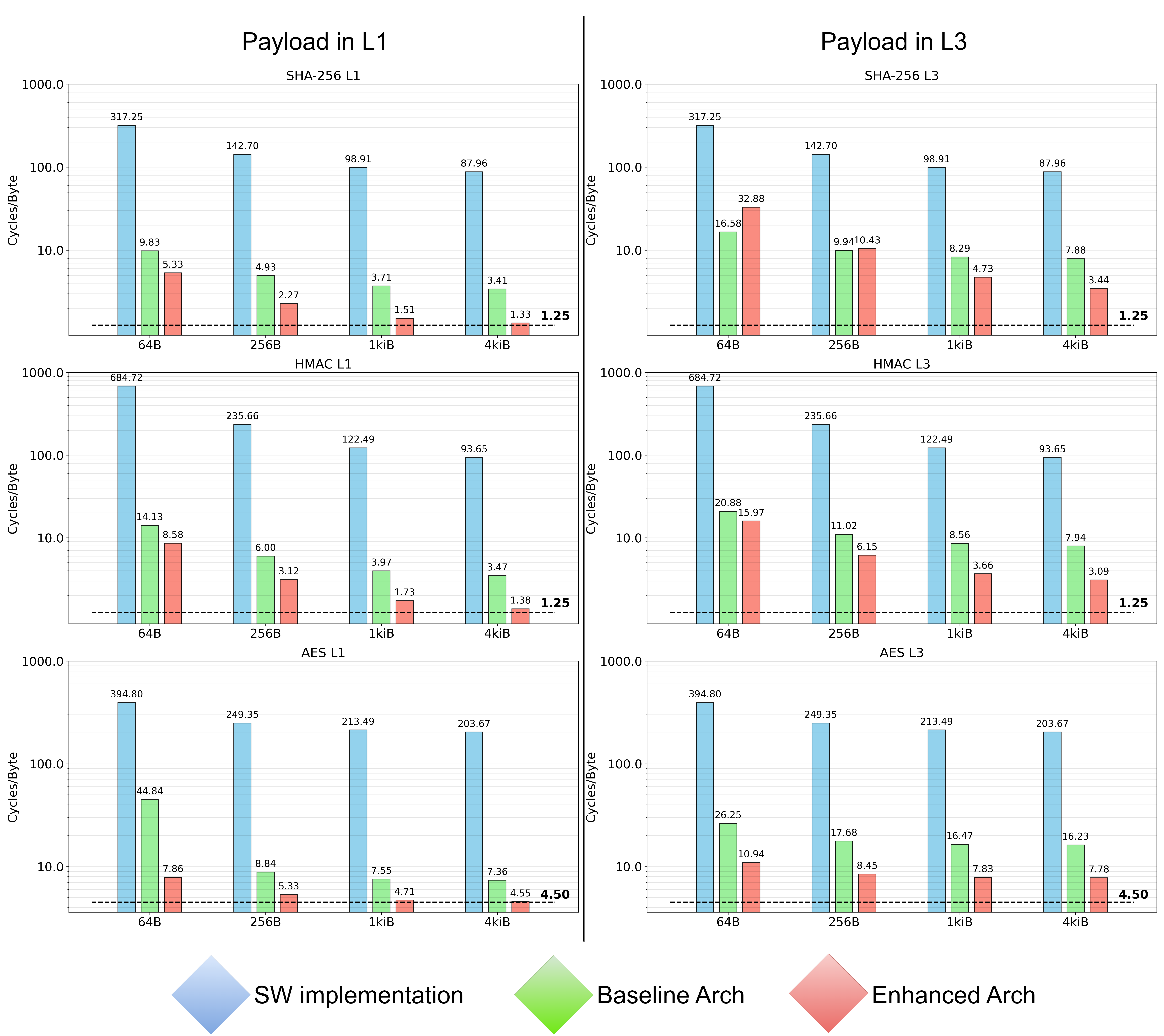}
  \caption{This figure shows the plots comparing the performance between CVA6 software execution, baseline architecture including the TLUL-to-AXI4 bridge, and the optimized architecture.}
  \label{fig:acceleratros}
\end{figure}

Figure~\ref{fig:acceleratros} presents a performance comparison, $Clks/B$, between the foundational architecture enhanced solely with the TLUL-to-AXI4 extension, already studied in \cite{parisi2024assessing}, and a more advanced architecture that incorporates DMA, TCDM, and improvements to the FIFOs of the TLUL interconnect. The former is represented with green bars and the latter with red bars. The experiments have been repeated with different payloads and for both the cases where the payload is stored into the external L3 memory (on the right side of the plot) and in OpenTitan's private scratchpad L1 memory (on the left side of the plot). In blue, the reference software implementation of each algorithm, where OpenTitan is not involved at all. The latter is the same experiment for each algorithms in both the plots on the right and on the left, as the software benchmarks run in one only configuration with payload in L3.
Across all benchmarks for payloads as small as 64 Bytes, a notably poor performance is observed, primarily due to the overhead associated with configuring the accelerators before and after task execution. 
The setup and tear-down processes overshadow the data transport and processing times for such small payloads. 
This effect is more pronounced as detailed in Table ~\ref{tab:opentitan_performance}, which breaks down the percentage of clock cycles dedicated to different phases for different payloads on the extended architecture, highlighting that 40\% to 50\% of the cycles are consumed by configuration and finalization stages for smaller payloads. 

When considering the enhanced architecture with payloads above 1KB, especially when utilizing the embedded scratchpad L1 memory, performance aligns more closely with the accelerators' theoretical bandwidth capabilities, as the actual task's processing dominates over the initialization and tear-down phases. In the plot on the left side where the payload is stored in OpenTitan's L1, the red bars that are associated to the enhanced architecture approaches the corresponding theoretical bandwidth provided by lowRISC documentation. Since in this configuration, the DMA and TCDM are not exploited as the payload is already stored within OpenTitan's L1.
Therefore, configuring the FIFOs in pass-through configuration within the TLUL interconnect is sufficient to fully utilize the available bandwidth of the accelerators for payloads stored in OpenTitan's L1. 
The comparison between the foundational architecture and the enhanced architecture, when the payload resides in L1, reveals that for HMAC and SHA-256, the performance speeds up by a factor of 2.7$x$. In contrast, the AES speeds up by a factor of 1.6$x$.
Under these conditions, the bandwidth of the accelerators is saturated as the performance metrics approach the nominal bandwidth.
This represents the maximum bandwidth the architecture can reach, underlining the maximum speed up reachable with respect to Parisi et al. results on the baseline lowRISC's architecture.
As we acted on the transport datapath, we observed a larger speed-up for SHA-256/HMAC because they are less computationally complex with respect to AES, being more memory-bound. The most significant speed-ups are observed on memory-bound tasks by enhancing the transport datapath.
The plot shows that considering a 4KB payload, the maximum speed up obtained on the extended architecture with respect to the reference software implementation is 67$x$ for SHA-256, 72$x$ for HMAC, and 44$x$ for AES.

When comparing scenarios utilizing TLUL-to-AXI4 and its datapath against those employing TCDM with DMA with the payload located into the L3 memory, on the right size of Figure \ref{fig:acceleratros}, a similar enhancement in speed is observed. However, the minimum bandwidth threshold is not met.
The DMA is issued twice, before and after processing the payload; thus, it does not implement double buffering techniques.
During task execution, the microcontroller accesses the embedded TCDM, presenting the same situation as the L1 case and reaching the same performance.
The benchmarks employed for the external L3  memory have been extended with instructions to issue the DMA twice.
For this reason, we expect similar performance as the L1 case where we saturate accelerators' bandwidth, but with some degradation introduced by the DMA, which is expected to be less than the degradation introduced by the TLUL-to-AXI4 datapath, requiring 23 clock cycles for each load/store.
The DMA overhead prevents the benchmark from reaching the same performance as the benchmark executed with the payload within the L1 memory.
The latency introduced by the DMA, measured at 1.4 $Clks/B$, is incurred twice for each benchmark accessing L3 memory to load the payload and store the result of the processing.
Despite this, the performance degradation in the L3 benchmarks compared to L1 is less than 2.8 $Clks/B$, which aligns with expectations since approximately 30\% of the DMA latency is obscured by IP configuration, resulting in about 2 $Clks/B$. The benchmarks' results are consistent with the sum of the nominal bandwidth of the accelerators and 70\% of the DMA's measured latency.
Moreover, as payload sizes increase, most clock cycles in L3 benchmarks are spent awaiting DMA data transport when considering SHA-256 and HMAC.
Overall, the performance we observe in the L3 case still presents significant speed up when the DMA and the TCDM are employed with respect to the baseline architecture.
When considering the base architecture with the TLUL-to-AXI4 bridge and its datapath, the performance is about 2$x$ worse with respect to the extended architecture.
This is expected as the TLUL-to-AXI4 without the interconnect optimizations introduces 23 clock cycles for each transaction.
Despite the overhead introduced by the DMA, even in large payloads, the overall clock cycle count is still less than employing the base architecture, showcasing the same speeds up between the two architectural configurations observed in the L1 benchmarks.
Integrating DMA and a dedicated TCDM markedly improves performance compared to a software implementation.
Moreover, the architectural extension based on DMA and TCDM speeds up the performance by 2.7$x$ when considering HMAC and SHA-256 and a speed-up of 1.6$x$ for AES.
By addressing and mitigating the critical bottleneck presented by the TLUL interconnect, we achieve full utilization of the accelerators' bandwidth within the OpenTitan framework with payloads larger than 1KB when working within the OpenTitan perimeter.

\section{Conclusion}
We presented a comprehensive integration and customization framework of the OpenTitan project to fit into a broader System-on-Chip environment. 
By successfully embedding the Earl Grey architecture into the reference SoC, we have demonstrated the feasibility and flexibility of OpenTitan as a silicon-ready root-of-trust.
By customizing memory hierarchies to align with available silicon area constraints and bypassing lifecycle state transactions, we have tailored the Earl Grey architecture to meet specific deployment needs without compromising its core security features. 
Establishing a mailbox-based secure communication channel, adhering to the ARM SCMI standard, facilitates secure and efficient task offloading from the host to the secure subsystem.
We adopted OpenTitan's deployment framework to implement our extensions to the architecture, extending it for the deployment in different hardware implementation targets such as FPGA emulation, ASIC, and RTL simulation, mapping OpenTitan's primitives to the target technology nodes.
We then optimized the Earl Grey datapath, customizing the TLUL interconnect and integrating a DMA engine with a dedicated TCDM to overcome performance bottlenecks when employing the cryptographic accelerators.
In conclusion, we implemented the reference SoC with the Earl Grey extended architecture in Global Foundries 22nm FD-SOI technology node, resulting in a total silicon area of 7.28$mm^2$, of which OpenTitan occupies 1.7$mm^2$.
We then benchmarked the enhanced architecture's performance, analyzing the latency introduced by the external memory hierarchic levels, presenting significant improvements in cryptographic processing speed, achieving up to 2.7$x$ speedup for SHA-256/HMAC and 1.6$x$ for AES accelerators, compared to the baseline Earl Grey architecture. The bandwidth requirements are met when considering payloads stored within OpenTitan perimeters, maximizing the accelerators' activity factor.

%%\input{sections/tables/results}

\begin{acks}
This work was supported in part by the UAE Technology Innovation Institute (TII), in part through the TRISTAN (101095947)
project that received funding from the HORIZON CHIPS-Joint Undertaking
programme, and in part by the Spoke 1 on Future High-Performance-
Computing (HPC) of the Italian Research Center on High-Performance
Computing, Big Data and Quantum Computing (ICSC) that received funding
from the Ministry of University and Research (MUR) for the Mission 4–Next
Generation EU programme.
\end{acks}

\bibliographystyle{ACM-Reference-Format}
\bibliography{main}

\end{document}